\documentclass[preprint2]{aastex}

\newcommand{\possessivecite}[1]{\citeauthor{#1}'s \citeyear{#1}}

\usepackage{pdfpages}
\usepackage{amsmath}

\shorttitle{The star formation in the center of 30 Doradus: the
  starburst cluster NGC~2070} \shortauthors{Cignoni et al.}

\begin{document}


\title{Hubble Tarantula Treasury Project: Unraveling Tarantula's
Web. II. Optical and Near Infrared Star Formation History of the Starburst
Cluster NGC 2070 in 30 Doradus}%

\author{
M. Cignoni\altaffilmark{2},
E. Sabbi\altaffilmark{2}, 
R. P. van der Marel\altaffilmark{2}, 
M. Tosi\altaffilmark{3},
D. Zaritsky\altaffilmark{4},
J. Anderson\altaffilmark{2}, 
D. J. Lennon\altaffilmark{5}, 
A. Aloisi\altaffilmark{2},
G. de Marchi\altaffilmark{6},
D. A. Gouliermis\altaffilmark{7},
E. K. Grebel\altaffilmark{8},
L. J. Smith\altaffilmark{9},
P. Zeidler\altaffilmark{8}
}
\email{cignoni@stsci.edu}

\altaffiltext{1}{Based on observations with the NASA/ESA Hubble Space Telescope, obtained at the Space Telescope Science Institute, which is operated by AURA Inc., under NASA contract NAS 5-26555}
\altaffiltext{2}{Space Telescope Science Institute, 3700 San Martin Drive, Baltimore, MD, 21218, USA }
\altaffiltext{3}{Istituto Nazionale di Astrofisica, Osservatorio Astronomico di Bologna, Via Ranzani 1, I-40127 Bologna, Italy}
\altaffiltext{4}{Steward Observatory, University of Arizona, 933 North Cherry Avenue, Tucson, AZ 85721, USA}
\altaffiltext{5}{ESA - European Space Astronomy Center, Apdo. de Correo 78, 28691Villanueva de la Ca\~{n}ada, Madrid, Spain}
\altaffiltext{6}{Space Science Department, European Space Agency, Keplerlaan 1, 2200 AG Noordwijk, The Netherlands}
\altaffiltext{7}{Universit\"at Heidelberg, Zentrum f\"ur Astronomie, Institut f\"ur Theoretische Astrophysik, Albert-Ueberle-Str.~2, 69120 Heidelberg, Germany}
\altaffiltext{8}{Astronomisches Rechen-Institut, Zentrum f\"ur Astronomie der
Universit\"at Heidelberg, M\"onchhofstr. 12-14, 69120 Heidelberg, Germany}
\altaffiltext{9}{ESA/STScI, 3700 San Martin Drive, Baltimore, MD, 21218, USA}

\begin{abstract}

We present a study of the recent star formation of 30 Doradus in the
Large Magellanic Cloud (LMC) using the panchromatic imaging survey
Hubble Tarantula Treasury Project (HTTP). In this paper we focus on
the stars within 20 pc of the center of the massive ionizing cluster
of 30 Doradus, NGC~2070. We recovered the star formation history by
comparing deep optical and NIR color-magnitude diagrams (CMDs) with
state-of-the-art synthetic CMDs generated with the latest PARSEC
models, which include all stellar phases from pre-main sequence to
post-main sequence. For the first time in this region we are able to
measure the star formation using intermediate and low mass stars
simultaneously. Our results suggest that NGC2070 experienced a
prolonged activity. In particular, we find that the star formation in
the region: i) exceeded the average LMC rate $\approx 20$ Myr ago; ii)
accelerated dramatically $\approx 7$ Myr ago; and iii) reached a peak
value 1-3 Myr ago. We did not find significant deviations from a
Kroupa initial mass function down to 0.5~M$_{\odot}$. The average
internal reddening E(B$-$V) is found to be between 0.3 and 0.4 mag.
 
\end{abstract}





\section{Introduction}

The Large Magellanic Cloud (LMC) harbors the nearest giant
extra\-galactic H\,{\sc ii} region \citep{kenni91}, the Tarantula
Nebula (30 Doradus). The central region of the nebula is dominated by
the cluster NGC~2070, a collective of several dense sub-clusters
(\citealt{walborn97,sabbi12}) whose light is dominated by massive OB
stars. The most prominent of these sub-clusters is the very central
super star cluster (SSC) R~136.

This region has a number of characteristics that make it
extraordinary. First of all, it displays an extreme rate of star
formation (SF), with one quarter of the total massive recent ($<\,10$
Myr) star formation in the LMC contained within 15$'$ from 30 Doradus
\citep{kenni91}. Moreover, with a stellar mass of at least $
2.2\,\times 10^4 \mathrm{M}_{\odot}$ concentrated within a radius of
4.7 pc \citep{hunter95}, this cluster can be classified as a
relatively low-mass clone of more distant starburst clusters, likely
building blocks of starburst galaxies.

Over the years, there have been debates over whether or not NGC~2070
is a young analog of old globular clusters. Recent arguments seem to
suggest a positive answer. Using photometry \cite{andersen09} found
that NGC2070's initial mass function (IMF) is Salpeter-like down to
$1\,\mathrm{M}_{\odot}$, giving the first direct evidence that
solar-like stars are formed in a starburst cluster. Using multi-epoch
spectroscopy of O-type stars, \cite{henault12} measured a velocity
dispersion between 4 and 5 km\,s$^{-1}$, close to the expected value
if it were in virial equilibrium. In this case the velocity dispersion
would be low enough to allow the cluster to stay bound.

Due to the proximity of the LMC, 30 Doradus can be imaged with the
Hubble Space Telescope (HST) on scales down to $\sim 0.01$ pc,
resolving stars down to the sub-solar regime. By coupling HST imaging
with ground-based spectroscopy diagnostics, we know that the region
has undergone continuous SF activity or a superposition of multiple
bursts of SF. Indeed, over the whole region, there is evidence of at
least three events: i) an on-going off-center activity, as documented
by the presence of embedded O-type stars and luminous infrared
protostars along an arc of molecular gas and warm dust around 30
Doradus \citep[]{rubio92,hyland92,walborn97,rubio98, walborn99,
  brandner01}; ii) a recent event, represented by the SSC R~136
itself, 2 Myr old or less, as inferred from spectroscopy of the O
stars (\citealt{massey98}), 3-4 Myr old from HST imaging
(\citealt{hunter95}); iii) an ``old'' event, documented by the
presence of Hodge 301, a 25 Myr old cluster (\citealt{grebel00})
located only 3 arcmin northwest of R~136.

However, most of high resolution studies are limited to a few bands
and cover only small patches of the 30 Doradus complex. The Hubble
Tarantula Treasury Project (HTTP; Sabbi et al. 2015, in prep.), an HST
survey covering a $\sim 14\arcmin \times 12\arcmin$ wide area around
30 Doradus at high resolution, fills this gap. This unique data-set,
which combines resolution and spatial coverage, can be exploited in
varied ways, including studies of massive SF, low mass SF, the IMF,
the reddening distribution, etc.. In our current paper, we take the
opportunity to explore the SF during the last 50 Myr for the central
40 pc of 30 Doradus, a.k.a. the starburst cluster NGC~2070, by
comparing the observational color-magnitude diagrams (CMDs) with
state-of-the-art synthetic CMDs. These simulations incorporate the
latest (V.1.2S) set of PAdova and TRieste Stellar Evolution Code
(PARSEC) isochrones (see \citealt{bressan12} and \citealt{Tan14}), the
first theoretical library to include homogeneously all stellar phases
from pre-main sequence (PMS) to post-main sequence for all masses
between $0.1$ and $350\,\mathrm{M}_{\odot}$. For the first time, we
derive the history of the region using low and intermediate mass
stars, either in the PMS or main sequence (MS) phases,
\emph{simultaneously}. In particular, we exploit magnitudes and colors
of the PMS turn ons (hereafter TOn; see Section \ref{tons}), i.e. the
CMD loci where the PMS phase joins to the MS. This ``hook'' has been
demonstrated to be particularly sensitive to age (see, e.g.,
\citealt{Stauffer80,Belikov98,Baume03,stolte04,mayne10,cignoni10b}). We
use both the optical F555W vs F555W$-$F775W and the near infrared
(NIR) F110W vs F110W$-$F160W CMDs of the HTTP sample, which offer
complementary advantages in terms of higher spatial resolution (the
optical) and lower reddening sensitivity (the NIR).

To recover the SF rate of NGC~2070 since the beginning of its
activity, there are three important factors that must be accounted
for: differential reddening, LMC field contamination, and stellar
crowding. The first causes the CMD features to appear redder and more
scattered, mimicking older populations and lowering the age
resolution. Field contamination mostly adds low mass stars to the
sample, mimicking much older populations and a steeper IMF. Finally,
stellar confusion affects completeness and, therefore, the reachable
look-back time. Moreover, unaccounted incompleteness can also mimic
mass segregation.

The paper is structured as follows. In Section 2 we briefly describe
the observations. Section 3 is dedicated to the physics of the TOn
clock. In Section 4 we identify the general properties of the stellar
populations in the HTTP data-set, with special emphasis on
NGC~2070. In Section 5 we perform artificial star tests, the only way
to take crowding into account, and we recover the SFH of NGC~2070
using the synthetic CMD approach.  Field contamination is also
carefully discussed. Results are presented in Section 6 and compared
to the literature in Section 7. Section 8 compares NGC~2070 with other
starburst clusters. Our conclusions (Section 9) close the paper.

Throughout the paper, for the sake of simplicity, we will refer to the
entire HTTP data-set as ``30 Doradus'', the starburst cluster as
NGC~2070 and its core as R~136.

\section{Observations and Photometric Reduction}

\begin{figure*}[!t]
\centering \includegraphics[width=16cm]{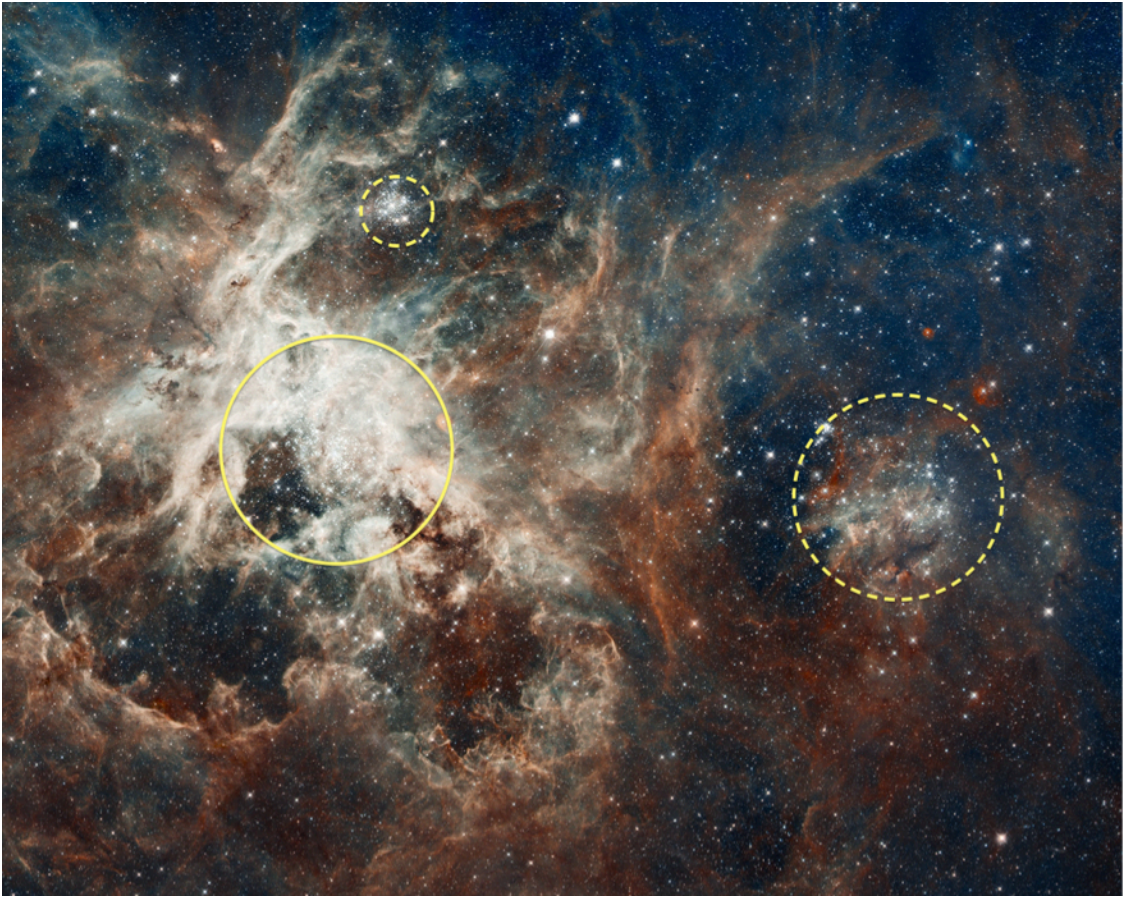}
\caption{F775W image of 30 Doradus. The stellar concentrations
  of NGC~2070 (continuous line cicrcle), Hodge~301 (dashed small circle) and
  NGC~2060 (dashed large circle) are also indicated.}
\label{30D_full} 
\end{figure*}

We observed 30 Doradus with the HST Wide-Field Camera 3 (WFC3) and the
HST Advanced Camera for Surveys (ACS) as part of proposal GO-12939
(PI: E. Sabbi). We built this program on an existing HST monochromatic
survey in the F775W filter (GO-12499, PI: Lennon), designed to measure
proper motions of runaway candidates. In both data-sets we used the
Wide Field Channel (WFC) of ACS in parallel with either the UVIS or
the IR channels of WFC3 to maximize the efficiency of the
observations. The images were taken with the filters F275W, F336W,
F555W, F658N, F110W and F160W between December 2012 and September
2013. The survey utilized 60 orbits. Figure \ref{30D_full} shows the
F775W image of 30 Doradus with indicated the major stellar
concentrations, NGC~2070, Hodge~301 and NGC~2060.

Bright and faint sources are respectively identified using the
packages {\tt img2xym\_WFC.09x10} (\citealt{anderson06}) and KS2, an
evolution of the program described in \cite{anderson08}. A detailed
description of the photometric analysis can be found in Sabbi et
al. (2015, in preparation).

We culled the catalog of detected objects to only include sources with
high quality in the PSF-fitting, Q$_{fit} > 0.75$. The final catalog
contains $\sim 30,000$ detected in the filter F275W, $\sim 100,000$ in F336W,
$\sim 400,000$ in F555W, $\sim 130,000$ in F658N, $\sim 620,000$ in
F775W, $520,000$ stars in F110W and $570,000$ in the F160W.  \\\\
\section{Stellar clocks: MS Turn-Off and PMS Turn-On}
\label{tons}
One of the characteristics of 30 Doradus that makes it a particularly
interesting object is the high concentration of massive stars that
coexist with a plethora of coeval intermediate and low mass PMS
stars. This enables us to reconstruct the past history of 30 Doradus
using the MS turn-off (MSTO)\footnote{Although the term MSTO refers to
  the end of the MS phase for stars of any mass, the reference here is
  to stars more massive than $8\,\mathrm{M}_{\odot}$, whose MS
  evolutionary times are shorter than $\approx\,50$ Myr.}, the locus
of the CMD where MS stars exhaust their core hydrogen, and the PMS
TOn, where low and intermediate mass PMS stars ignite hydrogen in
their cores. In terms of stellar mass, the MSTO mass is the mass of the
most massive star still on the MS at an evolutionary time
corresponding to the age of the cluster, while the TOn mass is the
mass of the least massive star that has reached the MS at the age of
the cluster. In the following we discuss pros and cons of the two
clocks.

\begin{figure*}[!t]
\centering \includegraphics[width=16cm]{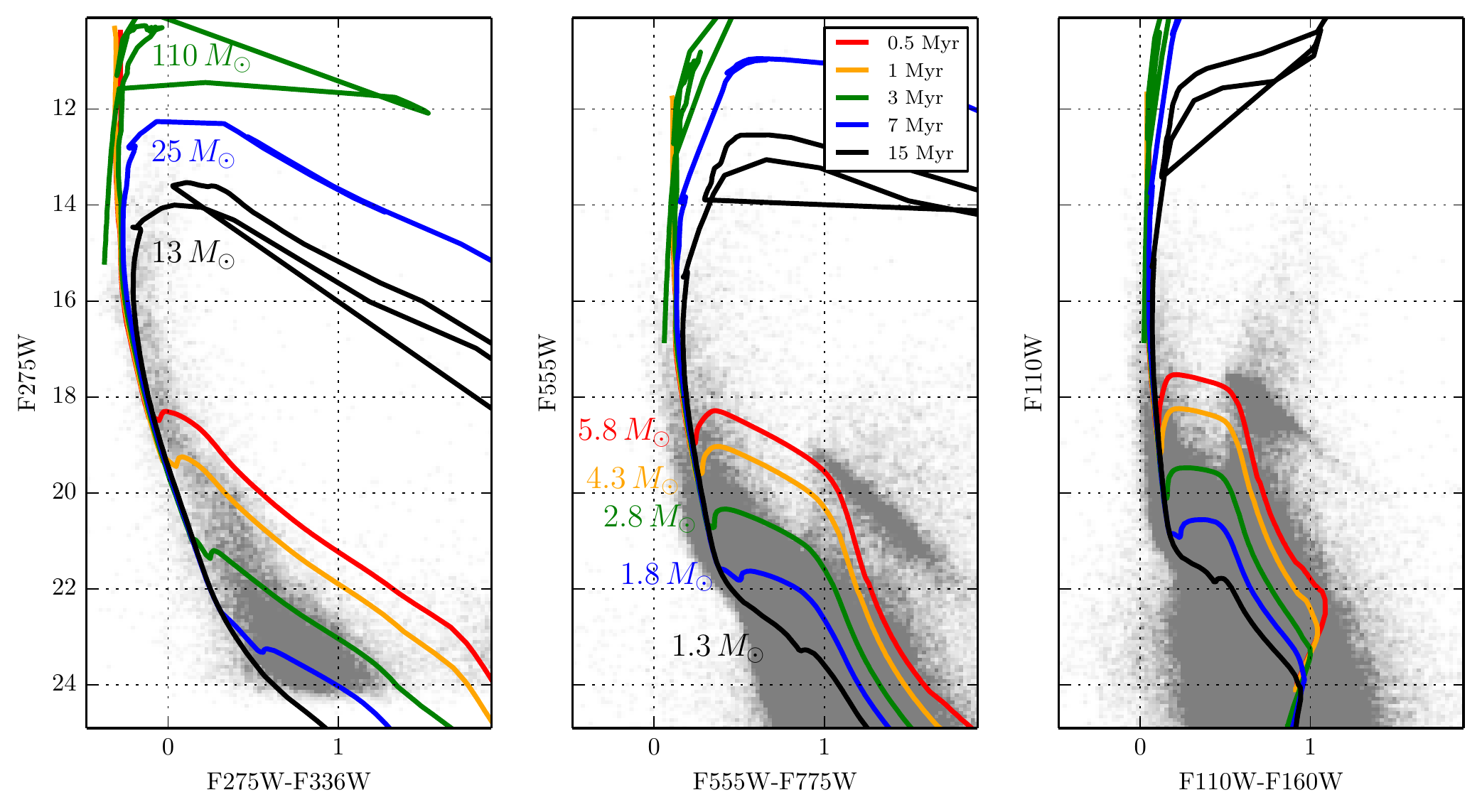}
\caption{Stellar isochrones of the labeled ages super-imposed on the
  entire HTTP data-set in the UV (left panel), optical (middle panel)
  and NIR (right panel) CMDs. MSTO and TOn masses (at the age of the
  corresponding isochrones) are also indicated in the UV and optical
  CMD, respectively.  The adopted distance modulus (m$-$M)$_{0}$ is
  18.5 and E(B$-$V)=0.3.}
\label{3iso} 
\end{figure*}

\emph{Massive stars:} The main appeal of using the MSTO of massive
stars (M$>8\,\mathrm{M}_{\odot}$) stems from the high luminosity of
such sources, which translates into high quality photometry and
complete samples. The optical and infrared spectra of these objects
are well approximated by the Rayleigh-Jeans tail of a black body with
temperature T$_{eff}$. Hence, given that the spectral energy
distribution shape is almost unchanged as a function of wavelength,
the optical and infrared colors are nearly constant (and around 0
mag). This is clearly visible in Figure \ref{3iso}, where we overlaid
young isochrones of different ages (0.5, 1, 3, 7, 15 Myr) on to the
entire HTTP catalog (shaded grey area) in the UV (left panel), optical
(middle panel) and NIR (right panel) CMDs. The adopted distance
modulus (m$-$M)$_{0}$ and reddening E(B$-$V) are 18.5
(\citealt{panagia91,schaefer08,pietr13}) and 0.3 (chosen by eye based
on the fit of the optical CMD), respectively. In the NIR CMD the upper
MS (UMS) looks like a vertical line and isochrones of different age
have MSTOs almost indistinguishable from one another. This degeneracy
is attenuated in the optical and almost lifted in the UV
CMD. Observationally, studies of massive stars has always been
hampered by their rarity, due to the short timescales involved in
their evolution and the steepness of the IMF.

On the theoretical side, models of massive stars are still affected by
major uncertainties. In particular, physical mechanisms like mass
loss, rotation and binary evolution are not well understood (see,
e.g., \citealt{demink12}).

\emph{Intermediate and low mass stars:} For ages younger than 20-30
Myr, the PMS TOn is another valuable stellar chronometer which
involves intermediate and low mass stars instead of massive
stars\footnote{Stars more massive than 6 M$_{\odot}$ have no PMS phase
  at all.}. In analogy with the MSTO, the TOn properties are directly
related to the age of the stellar population, but with evolutionary
times much shorter than the corresponding MS times. In fact, the age
of a cluster is equal to the time spent in PMS phase by its most
massive star still in PMS phase. By definition, this star is at the
TOn. Hence, when the intrinsic luminosity of the TOn is detected, it
is straightforward to associate it with the age of the cluster. From
the CMD point of view, the potential strength of the TOn is apparent
from the morphology of the isochrones in Fig. \ref{3iso}. In the
optical and NIR CMDs the isochrone portion just before the MS has a
hook and then is significantly flatter than the MS. The TOn is at the
vertex of the hook, quite easy to recognize. For ages older than
20$-$30 Myr the PMS phase is much closer to the MS and the TOn
visibility declines. Theory also predicts that the PMS phase
(recognizable as the portion of the isochrones at the right of the MS
in Fig. \ref{3iso}) itself is a valuable age indicator: older PMS
isochrones are fainter and closer to the MS than younger
ones. However, poorly understood phenomenons like residual mass
accretion and magnetic fields (whose interplay is responsible for the
appearing of irregular photometric variability and UV to infrared
excesses), and observational uncertainties like differential reddening
from the circumstellar material can dislocate these stars (especially
in the first few Myr) from their theoretical positions in the CMD (see
\citealt{goul12} for a review).

An advantage of the TOn with respect to the MSTO derives from the
evolution at nearly constant luminosity of PMS stars near the MS,
corresponding to an almost horizontal track in the optical and NIR
CMDs. This leads to a luminosity(TOn) - age relation. On the other
hand, the MS evolution of massive stars near the MSTO is rather vertical
(except in the UV CMD), hence age is not uniquely related to
luminosity. Among the drawbacks, the intrinsic faintness of TOns
compared to the MSTOs makes TOns prone to photometric errors and
incompleteness, issues that are exacerbated in the UV CMD because
older TOns tend to be also redder. From the theoretical side, the
TOn visibility is intimately connected with the PMS evolutionary times
which are still model dependent (see, e.g.,
\citealt{baraffe09,hoso011,soderblom14} and references therein).

In this paper we aim to study the SFH of 30 Doradus with the TOns,
hence we focus our analysis on the optical and NIR CMDs. As shown in
the optical CMD (middle panel of Fig. \ref{3iso}), the most massive
star that is relevant to this study is around $6\,\mathrm{M}_{\odot}$.

Before closing this section, we emphasize another interesting feature
that emerges from Fig. \ref{3iso}. Whereas the isochrones reddened
with E(B$-$V)= 0.3 fit well the UMS in the optical and NIR CMDs
(middle and right panel), the same isochrones are clearly too blue in
the UV CMD (left panel). The explanation for this effect resides in
the different populations that these CMDs trace: most of the UV stars
are very young 30 Doradus stars, with minimal contamination from field
stars (the RC is hardly visible at F275W$\sim 23$), while the optical
and NIR stars can be both members and field stars. As a consequence,
the average reddening of the young population in 30 Doradus must be
\emph{higher} than the adopted value E(B$-$V)= 0.3, while the average
reddening of the field can be as low as the foreground MW reddening
(E(B$-$V)= 0.07; \citealt{FS84}).

The next Section is dedicated to a cursory CMD inspection of the
stellar populations in the whole HTTP data-set.  \\\\

\section{Stellar populations in the HTTP catalog}

The CMD is the most powerful tool to recover the history of a resolved
stellar population, since different parts of the CMD are populated by
different masses with different evolutionary times. In the following
analysis we use the CMD to investigate which populations are present
in the whole HTTP catalog and the role of the NGC~2070 region. For
this task we use the optical CMD, which offers better spatial
resolution than the NIR one. A more quantitative analysis will be the
subject of Section \ref{SFR}.

\subsection{The whole HTTP sample}

\begin{figure*}[!t]
\centering \includegraphics[width=12cm]{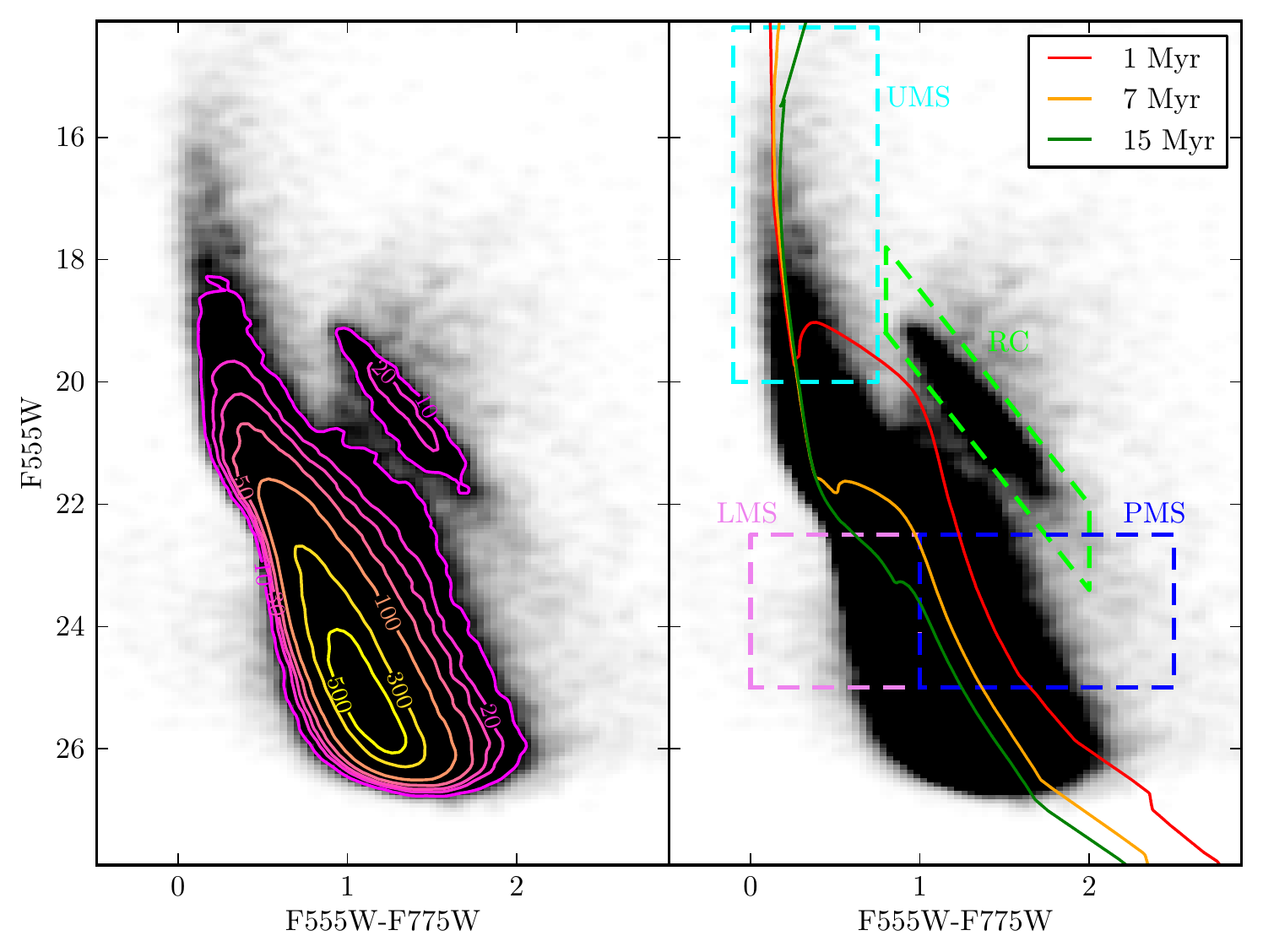}
\caption{Optical CMD of all stars in the HTTP catalog with overlaid
  isodensity contours (left panel) and PARSEC isochrones for the
  labeled ages (distance modulus (m$-$M)$_{0}$ and reddening E(B$-$V)
  are 18.5 and 0.3, respectively). The dashed selections indicate
  sample of UMS, LMS, PMS and RC stars (see text).}
\label{CMD_TOT} 
\end{figure*}

The F555W vs F555W$-$F775W CMD of the entire HTTP data-set is shown in
Fig. \ref{CMD_TOT}, with overlaid isodensity contours (left panel) and
stellar isochrones (PARSEC) of different ages (right panel).
The first remarkable feature of this CMD is an extended UMS, populated
by a plethora of intermediate and high mass stars. As the overlaid
isochrones show, the width of the UMS is hardly explained by a
difference in age. Moreover, given the average youth of these stars,
plausibly much younger than $50$ Myr, a metallicity spread is
unlikely. As widely discussed in other works (see, e.g.,
\citealt{selman99,demarchi14b} ), differential reddening is the main
cause of this effect, although stellar rotation may also have a role
in widening the UMS.

To the right of the UMS, the next striking feature is the very
elongated red clump (RC; see also Section \ref{rc_sec}) and a broad
red giant branch (RGB). These phases are populated by stars older than
1 Gyr and belonging to the general field population of the LMC. As for
the UMS width, most of the elongation and broadening is due to severe
differential reddening affecting the region (see
\citealt{has11,demarchi14a,demarchi14b}).

Finally, the lower MS shows a huge color dispersion (more than 1 mag)
at relatively bright magnitudes (F555W$\sim 23$). As usual, a possible
explanation comes from the comparison with the isochrones. Although
differential reddening contributes to this effect, pushing part of the
lower MS (LMS) to the red, the existence of short-lived massive stars
requires that a large fraction of these red stars are genuine PMS
stars (coeval with the massive stars).

More information on the nature of these populations can be inferred
from their spatial distribution. In fact, the spatial distribution of
stars in different evolutionary stages yields important information on
the star formation processes across the region. Assuming a velocity
dispersion of 21-27 km\,s$^{-1}$ (measured from RGB stars;
\citealt{carrera11}), stars older than 1 Gyr are expected to be
diffused over scale lengths of several kpc, hence their distribution
should be rather uniform over the HTTP field of view (FOV; $\sim$200
pc). On the other hand, stars like those in NGC~2070, young and with
very low velocity dispersion (4$-$5 km\,s$^{-1}$; \citealt{henault12}), are
expected to be close to their birthplace.

\begin{figure*}[!t]
\centering \includegraphics[width=12.5cm]{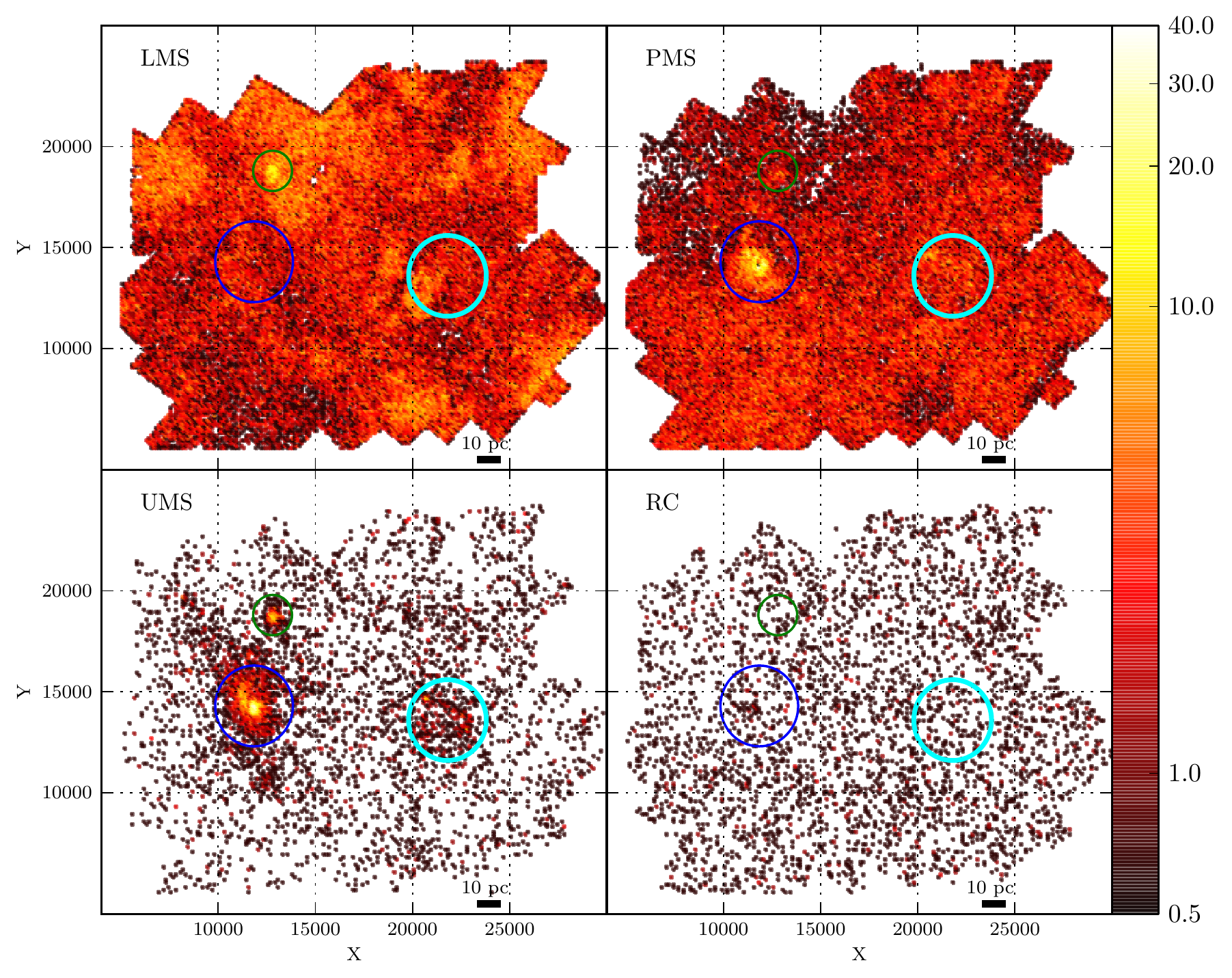}
\caption{Number of stars per pc$^2$ (see the color-bar on the right in
  logarithmic units) for different group of stars (see selection in
  Fig. \ref{CMD_TOT}): LMS (top-right panel), PMS (top-right panel),
  UMS (bottom-left panel), RC (bottom-right panel) stars. Blue, green
  and cyan circles highlight the regions of NGC~2070, Hodge~301 and
  NGC~2060, respectively.}
\label{ALL_XY} 
\end{figure*}

The right panel of Fig. \ref{CMD_TOT} highlights CMD regions of
selected UMS (cyan box), RC (green box), low mass MS (LMS) (pink box)
and PMS (blue box) stars. UMS and PMS stars are young objects a few
tens of Myr old, RC stars are intermediate age stars ($0.7-2$ Gyr),
LMS are MS stars older than few tens of Myr. Figure \ref{ALL_XY} shows
the corresponding spatial density (stars per pc$^2$).
Populations grow more compact as one moves toward younger ages (see
also Fig. 5 and 6 in \citealt{harris99}). Other interesting features
are:

1) the UMS stars (bottom left panel in Fig. \ref{ALL_XY}) appear very
clustered. Three major concentrations are visible: NGC~2070, the most
prominent, about 40 pc wide (encircled in blue), Hodge~301 (encircled
in green) and NGC~2060 (encircled in cyan). Lighter over-densities are
also visible throughout the FOV. In other words, although most of the
ongoing and recent (i.e., in the last 50 Myr) SF is concentrated in
few dense loci, a minor recent activity is present in the entire area;

2) as compared to the UMS distribution, RC distribution (bottom-right
panel) is, as expected, rather uniform. RC stars are a pure sample of
field stars. Their age, older than 700 Myr, rules out that these
objects are associated with 30 Doradus, which is, at most, a few Myr
old. The small over-densities are probably contamination of
intermediate mass PMS stars or very reddened massive stars from the
youngest regions. From a theoretical point of view, the intrinsic
position of the RC in the CMD could be affected by factors like
binaries, differences in age and metallicity, etc. (see, e.g.,
\citealt{castellani00}). However, these effects are insufficient to
account for the observed RC elongation (up to 1.5 mag in
F555W$-$F775W), which is mostly due to differential reddening. The
source of this reddening is the gas/dust located between the closest
and farthest RC star along the line of sight. More specifically,
background RC stars suffer the highest degree of absorption, which is
caused by the combination of Milky Way (MW) and 30 Doradus extinction,
while foreground stars will suffer only from MW (E(B$-$V)$\sim 0.07$)
extinction;

3) The distributions of LMS (top-left panel) and PMS (top-right panel)
stars appear mutually exclusive: the top-left corner of the FOV shows
a paucity of PMS stars, while LMS stars are clearly overabundant
there. The main cause for this effect is reddening. Most of the LMS
stars belong to the LMC field. This is because 30 Doradus is younger
than few Myr, hence its MS stars are necessarily brighter than our LMS
box of Fig. \ref{CMD_TOT} (right panel) (at ages $< 10$ Myr the 30
Doradus low-mass stars are mainly still in PMS, as suggested by the
isochrones in Fig. \ref{CMD_TOT}). Under these circumstances, the only
way to remove LMS stars from the LMS box is the action of reddening
(see also the map of Fig. \ref{RC_SPREAD}). Doing so, these reddened
MS stars are likely to fill the PMS box, eventually producing the
observed anti-correlation LMS/PMS stars (a similar behavior is found
by \citealt{goul06} in NGC~346; see their Fig. 4). In addition to
this, the tendency of young PMS stars to be concentrated where the
optical depth is higher exacerbates this effect. The only exceptions
are the center of NGC~2070, where the LMS stars are missed because of
the severe incompleteness, and Hodge~301, whose high concentration of
LMS is due to its higher age (so Hodge~301's TOn gets into the LMS
box).

Finally, the blue circle in Fig. \ref{ALL_XY} is the region we have
used to recover the SFH of NGC~2070. It is immediately clear from the
maps that the region harbors most of the UMS stars of the entire 30
Doradus complex, as well as a remarkable concentration of PMS
stars. On the other hand the region is quite deficient in LMS stars,
probably lost because of the extreme crowding conditions.


\subsection{NGC~2070}

\begin{figure*}[!t]
\centering \includegraphics[width=12cm]{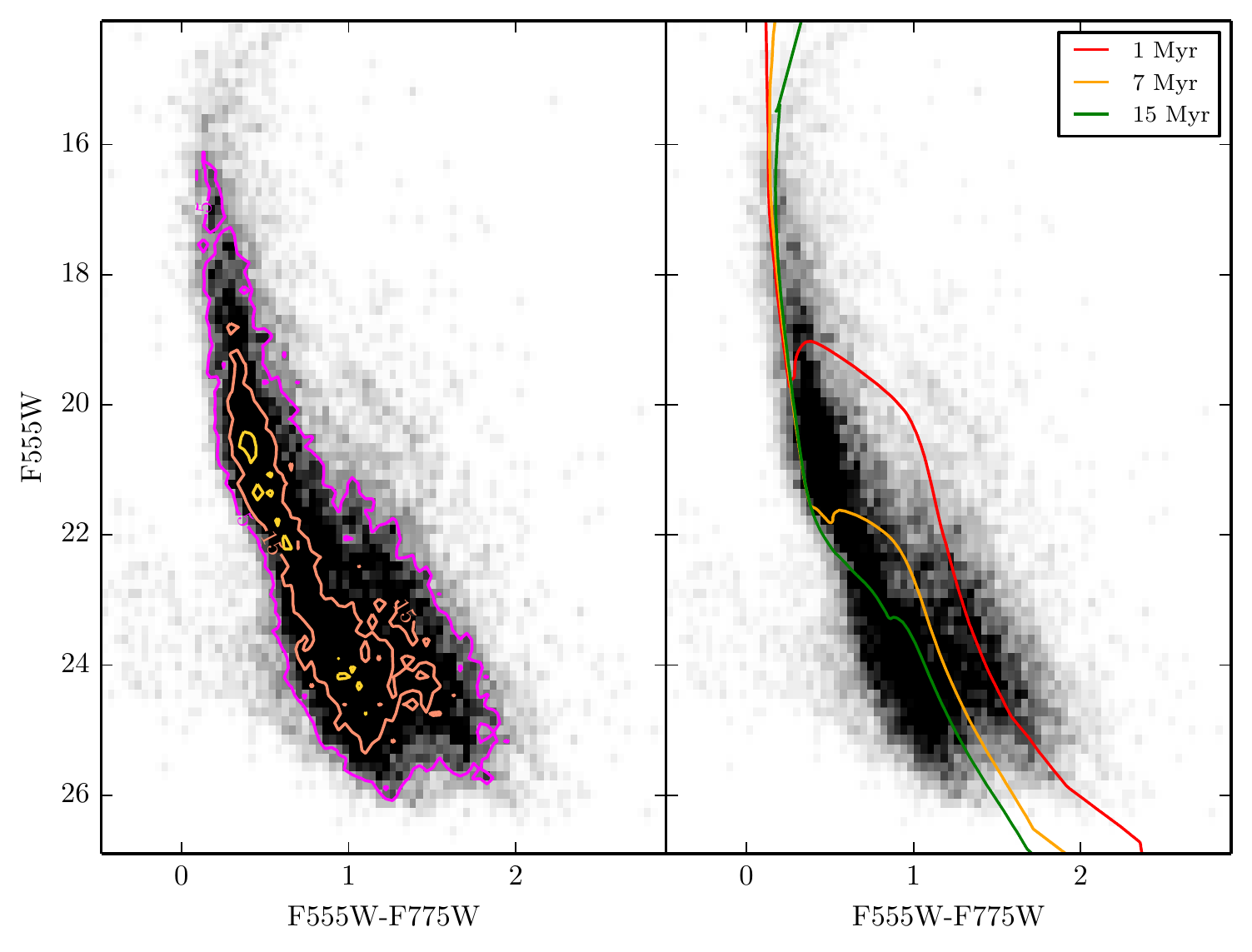}
\caption{CMD of NGC~2070 with isodensities overlaid (left panel) and
  PARSEC isochrones of 1, 7, 15 Myr overlaid (right panel). The
  adopted distance modulus (m$-$M)$_{0}$ is 18.5 and E(B$-$V)=0.3.}
\label{CMD_C} 
\end{figure*}

In this Section we discuss the broad CMD features of the stellar
population of NGC~2070. Our selection includes all stars within 2000
pixels ($\approx$ 20 pc) from the cluster center\footnote{Throughout
  the paper we will refer to R~136 as the NGC~2070 center because it is
  spatially well defined.}. Figure \ref{CMD_C} shows the corresponding
CMD for the F555W vs F555W$-$F775W filters with overlaid isodensity
contours (left panel) and PARSEC isochrones of 1, 7 and 15 Myr (right
panel). The MS contours of NGC~2070 differs from those of 30 Doradus
as a whole (Fig. \ref{CMD_TOT}). The peak density is around
F555W$\approx 21$, with minor peaks down to F555W$\approx 22$, while
the 30 Doradus CMD shows a smooth profile down to F555W$\approx 24$.
\begin{figure*}[!t]
\centering \includegraphics[width=10cm]{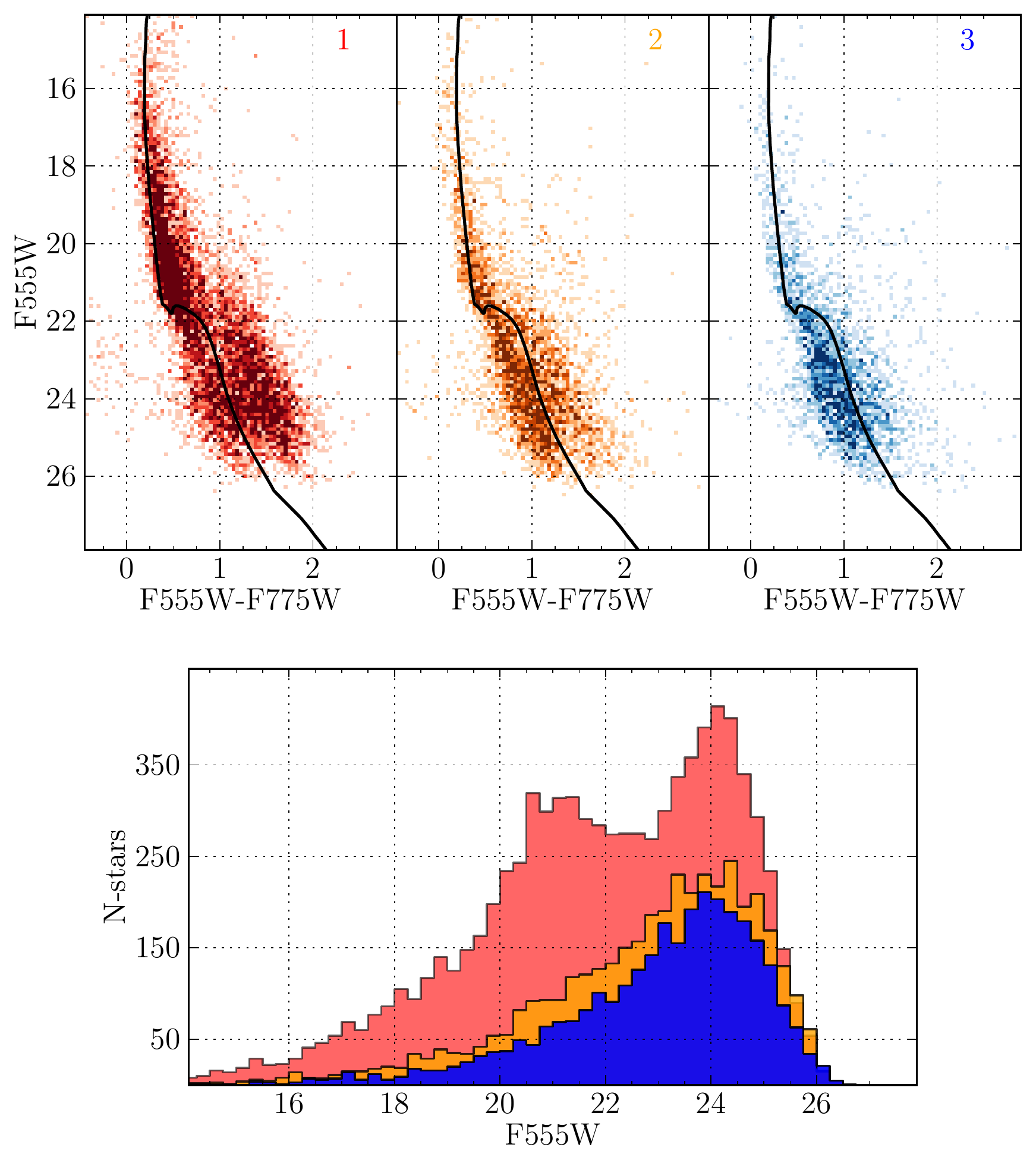}
\caption{NGC~2070 radial distribution. Top panels: from left to right,
  stars from progressively more external annular regions of equal area centered
  onto R~136. A 7 Myr old isochrone is also shown to guide the
  eye. Lower panel: the corresponding LFs. }
\label{CMD_C_LF_radial} 
\end{figure*}

In Figure \ref{CMD_C_LF_radial} we investigate the radial distribution
of stars in NGC~2070. The top panels show CMDs of stars in concentric
annuli (1-2-3) of equal area (the radius of the innermost circle is
about 12 pc) around the center of the cluster (from left to right,
progressively further out from the center), the bottom panel shows the
corresponding luminosity functions (LFs). We find that:

1) The LF of the innermost region 1 shows two clear peaks, located
at F555W$\approx 21$ and F555W$\approx 24$. The bright peak is well
fitted by an isochrone of 7 Myr (see the figure), evidence of a young
TOn. Stars brighter than this magnitude are mostly MS cluster members,
while fainter stars are members only if they are on the PMS (by the
definition of TOn, fainter members have not reached the MS yet).
Indeed, a visual inspection of the CMD shows a clear color bimodality
below F555W$\approx 21$, which is likely due to field MS stars (mostly
not-members) on the blue side, and PMS stars (members) on the red
side. The dip after the bright peak is caused by the short
evolutionary timescale of the PMS phase compared to the MS. After the
dip the LF rises again following the IMF, which increases at lower
masses. Eventually the incompleteness wins, creating the second
decline at F555W$\approx 24$. The apparent lack of lower MS stars in
the region 1 is probably due to the higher crowding, which causes more
severe incompleteness than in the more external annuli 2 and 3. In
Section \ref{SFR} we will use the synthetic CMD approach, combined
with artificial star experiments, to test if these lower MS stars are
compatible with LMC field contamination or hide some older TOns. To
this aim, we also need to estimate the LMC contamination in a
reference field (see next Section).

2) In contrast to the innermost region 1, annular regions 2 and 3 show
a monotonic increase towards fainter magnitudes, with no intermediate
peak before the final drop due to incompleteness. In terms of age, the
lack of obvious TOns means that regions 2 and 3 are not dominated by
as young stars like in region 1. Nonetheless, at the magnitude of the
region 1 peak, region 2's LF has a mild excess of stars compared to
region 3, which would point to a poorly populated TOn. As a further
proof, the region 2 CMD also shows an excess of PMS stars relative to
region 3.

3) Stars brighter than $V\sim 18$ in the region 3 show a larger color
spread than in regions 1 and 2. The RC is clearly visible in regions 1
and 2 (at magnitudes 20.0$-$20.3 and colors 1$-$1.5), while it is much
more dispersed in 3. All of these suggest a differential reddening
that is higher in region 3 than in regions 1 and 2.

\section{Recovering the SFH of NGC~2070}
\label{SFR}

The technique of recovering the SFH of spatially resolved populations
from their CMDs (e.g. \citealt{tosi91}) has undergone continuous
refinement and can now provide reliable SFHs for any resolved
population within $\approx 20$ Mpc (see, e.g.,
\citealt{tolstoy09,cignoni10} and references therein). These advances
have been achieved because of improved models of stellar evolution,
which are now computed for fine grids of stellar masses and
metallicities and faster multi-CPU computing facilities, which allow
one to perform extensive artificial star tests on the real images and
to fully explore a wide parameter space.

A widely used approach consists of populating a 2D array of basic
synthetic CMDs generated from stellar models. Each basic CMD is a
``fuzzy'' isochrone, with duration $\Delta t$, fixed metallicity and
an assumed IMF. To compare the basic CMDs and the observational
counterparts, the models are convolved with photometric errors and
incompleteness as derived from artificial star tests performed on the
real images. The best superposition of the basic CMDs defines the SFH
and the age-metallicity relation, and is the one that minimizes the
residuals from observational CMD. The approach adopted here is
described in more details in the Appendix.

Despite the advances in this field, the derivation of the SFH from the
CMD is often affected by systematic errors that are difficult to
assess. From a theoretical point of view, several stellar phases are
still uncertain (thermally pulsing asymptotic giant branch, PMS, and
post-MS for massive stars), while observationally differential
reddening and highly variable incompleteness can be hard to treat.

Concerning NGC~2070, and the general 30 Doradus region, we face three
major problems in trying to estimate the SFH: 1) membership errors due
to field interlopers from the LMC field that mimic older populations;
2) the extreme crowding conditions exacerbate the incompleteness,
shortening the reachable look-back time; 3) the high level of
differential reddening spreads and dims the CMD, blending together
young PMS stars and older MS stars, which introduces further age
ambiguities.


\subsection{Field contamination}

\label{rc_sec}

\begin{figure*}[!t]
\centering \includegraphics[width=14cm]{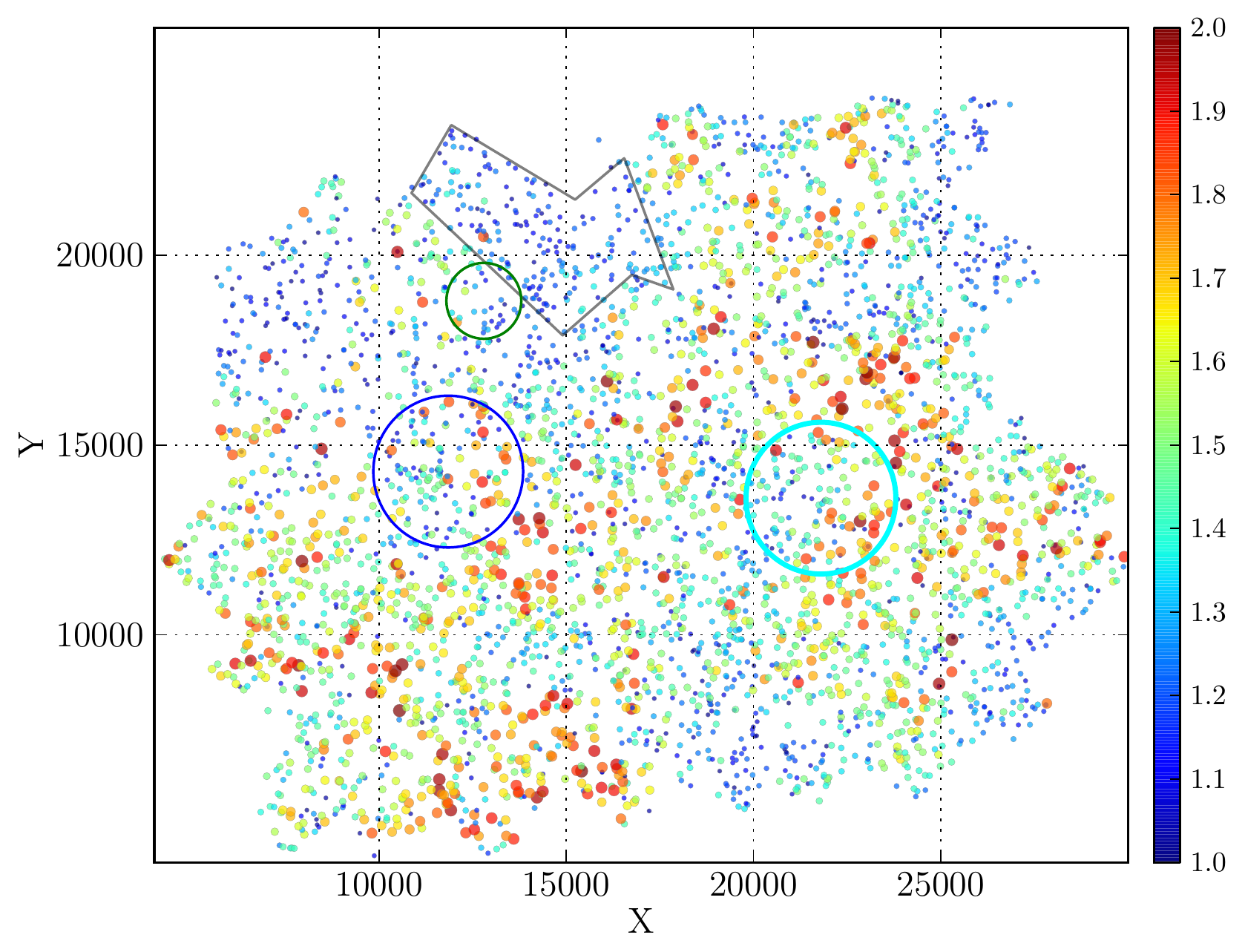}
\caption{Spatial distribution of RC stars color coded according to
  their F555W$-$F775W color as given on the vertical bar on the
  right. The area delineated by a black line represents a region with
  low extinction (see text).  Blue, green and cyan circles highlight
  the regions of NGC~2070, Hodge~301 and NGC~2060, respectively.}
\label{RC_SPREAD} 
\end{figure*}

To measure the SFH of NGC~2070 we need to estimate the local field
contamination. A typical approach is to decontaminate the cluster by
subtracting the star-counts from a reference field. However, given
reddening and incompleteness variations across the whole 30 Doradus,
it is impossible to find another direction replicating the
observational conditions of NGC~2070. A way out is to find a field
that resembles as much as possible the LMC field \emph{we would
  observe without} NGC~2070, which means low or negligible SF activity
in the last 50 Myr, minimal differential reddening and high
completeness down to F555W$\sim 24$. This reference LMC field could be
then artificially corrected for the more severe incompleteness and
photometric errors of NGC~2070. The resulting field would differ only
in terms of normalization and differential reddening from the real LMC
field contaminating NGC~2070. Finally, reddening and normalization of
this field could be tuned together with the SFH of NGC~2070 until an
adequate match of NGC~2070's CMD is found.

As a first step we searched for low reddening areas around 30 Doradus. A
good tracer for extinction is the RC color. Figure \ref{RC_SPREAD}
shows the spatial distribution of RC stars, color coded according to
the F555W$-$F775W color (larger and redder symbols correspond to
redder F555W$-$F775W). We find that, at odds with the smooth
distribution of the entire RC sample (see the bottom-right panel of
Fig. \ref{ALL_XY}), redder RC stars are very concentrated along
filaments and arcs, most likely tracing gas and dust in 30
Doradus. This allows us to select extended regions, like the wide
region just above NGC~2070 (see the black box in the Figure), with
relatively blue RC stars. Because these stars can not all be in the
foreground, these regions must be low-extinction windows in the
gas/dust layers of 30 Doradus. Interestingly, the aforementioned
region is also poorly populated by PMS stars (see top-right panel of
Fig. \ref{ALL_XY}), which implies minimal SF activity in the last 50
Myr.

\begin{figure*}[!t]
\centering \includegraphics[width=8cm]{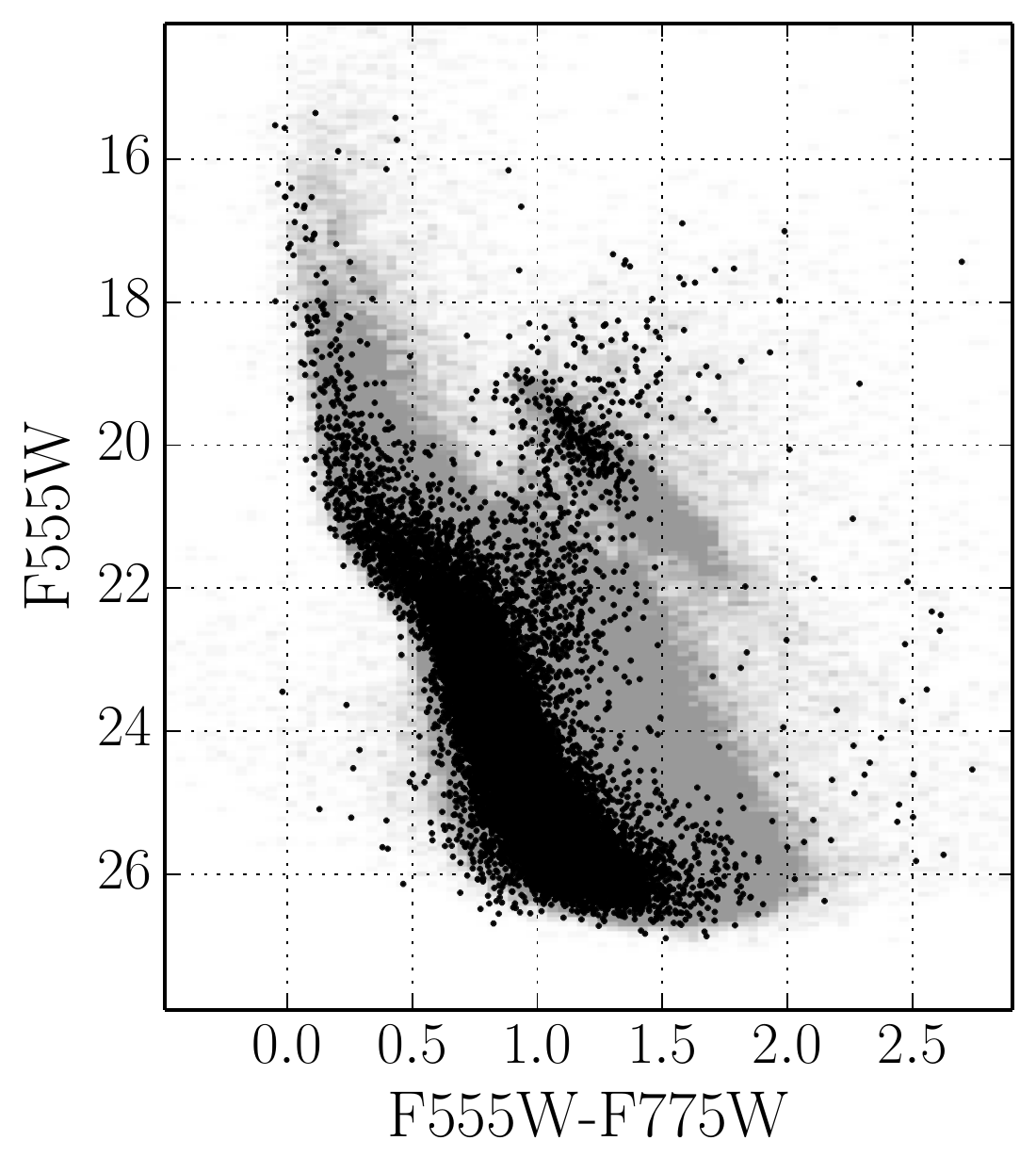}
\caption{CMD for the reference field (black dots) overlaid on the
  entire 30 Doradus sample (grey dots).}
\label{CMD_LOW_RED} 
\end{figure*}

After a careful inspection of all regions with low extinction and PMS
number, we found the best compromise in the region delineated by a
black line in Fig. \ref{RC_SPREAD}. The corresponding CMD is shown in
Fig. \ref{CMD_LOW_RED}, here overlaid on the 30 Doradus CMD (grey
symbols). This sample was adopted to represent our reference field. As
expected, its RC and lower MS are much tighter than the general CMD,
signatures of scarce differential reddening and young SF activity,
respectively. It is worth noting, however, that residual differential
reddening is still present.

\subsection{Artificial star tests}
\label{artifix}
To test the level of completeness of our photometric data and to have
reliable estimates of photometric errors, we ran extensive artificial
star experiments. The experiments consist of adding ``fake'' sources
for each of the eight pass-bands, modeled with the PSF used in the
photometric analysis of the frames, onto the actual images. We then
applied the same source detection routines used for our science images
to the fields containing the combined actual images and the fake
sources. We then determined the completeness fractions, defined as the
ratio of recovered artificial stars to the number of the injected
ones. We considered an artificial star lost if it is not recovered or
if it is recovered being 0.75 mag brighter than its input
magnitude. In fact, this means that it has fallen either on a real
star brighter than the artificial star or one of its same
brightness. Thus, we are not recovering or measuring the artificial
star but a real one instead.

To preserve the crowding conditions of the data, fake stars were
arranged in a spatial grid such that the separation of the centers in
each star pair was larger than two PSF radii. We found that a
separation of 20 pixels guarantees that the probability of recovering
a fake star is the same as it would be if we added only one fake
star. The experiment is then repeated using a series of slightly
shifted grids.

\begin{figure*}[!t]
\centering \includegraphics[width=11cm]{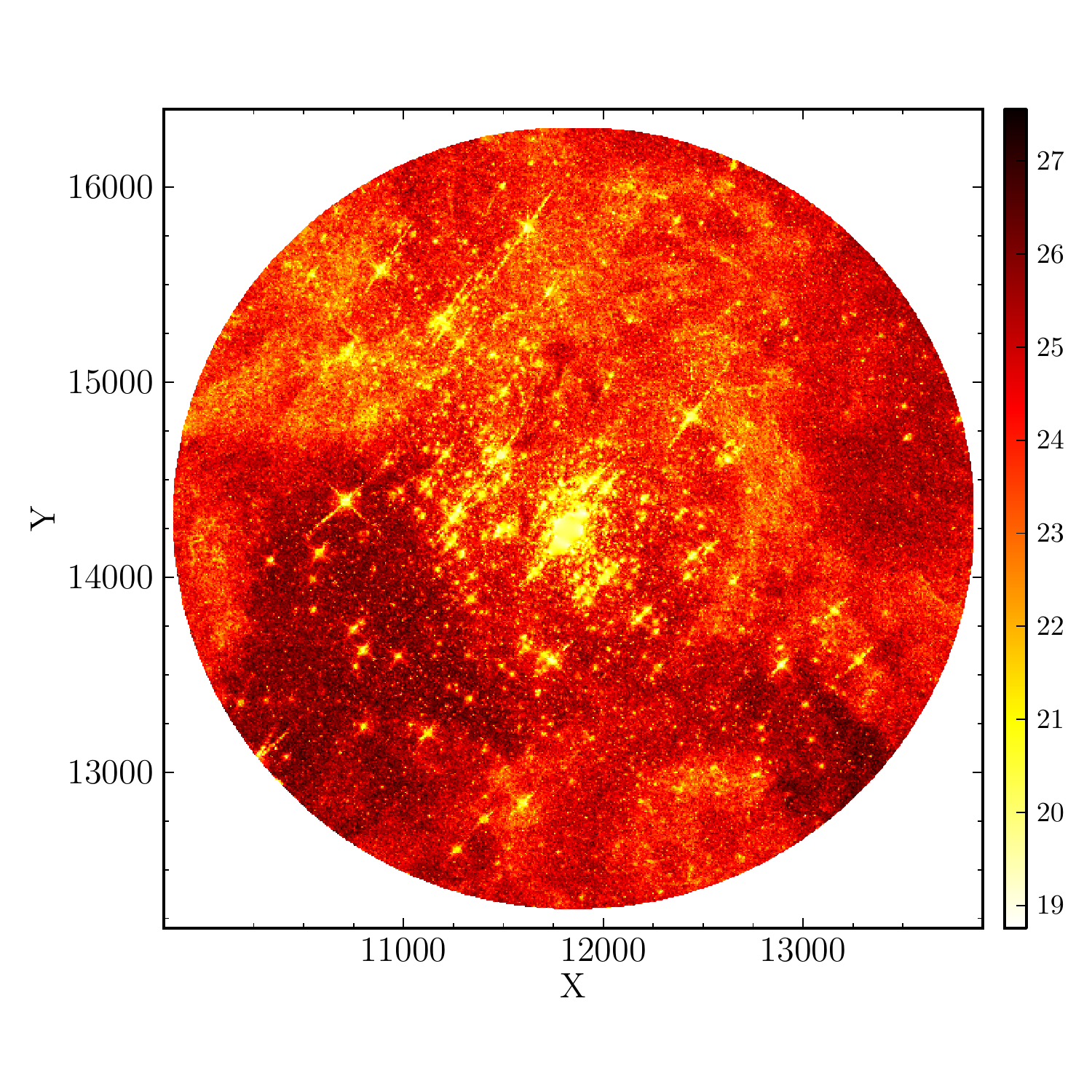}\\
\centering \includegraphics[width=8cm]{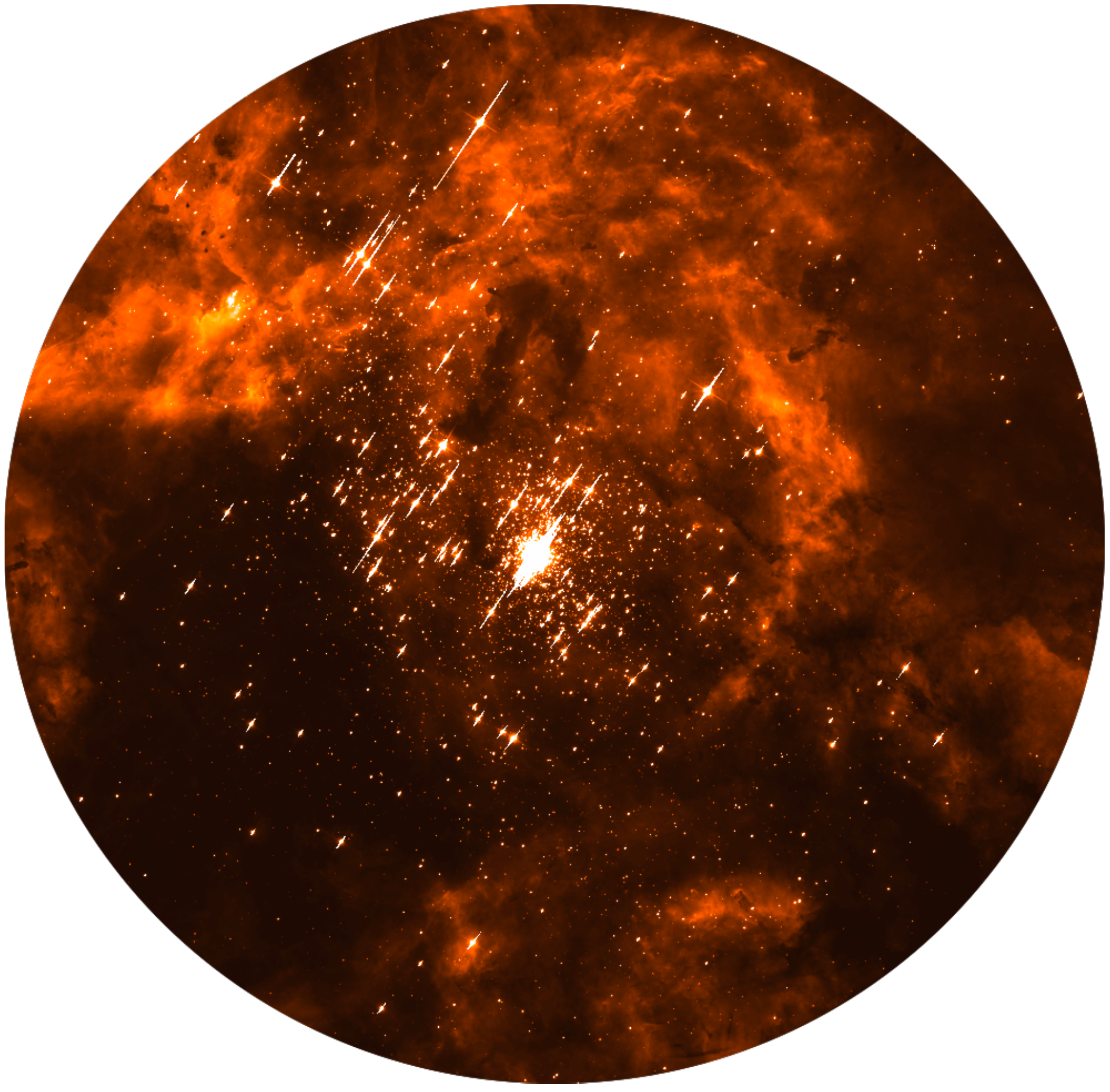}
\caption{Top panel: Completeness map for NGC~2070 color-coded
  according to the F555W magnitude where the sample is 50\% complete
  (which means that 50\% of the injected artificial stars are
  recovered). The highly incomplete central region is the core of
  NGC~2070, R~136. Bottom panel: the real image of the region in the
  filter F555W.}
\label{map50} 
\end{figure*}

As an example of our fake stars procedure, the top panel of Figure
\ref{map50} shows the impressive 50\% completeness map in the F555W
band for a region 2000 pixel ($\approx$ 20 pc) away from the cluster
center (5 million fake stars). The bottom panel shows the
corresponding real image. The most striking feature is the remarkable
completeness variation, up to 8 magnitudes, in different locations
within the region. Regions like the broad area around X=10750, Y=13500
are photometrically complete at 50\% level down to $V \approx 26$,
whereas more crowded sub-regions, like the central area, are 50\%
complete only for $V<20$.
Such variations are determined by the interplay of four major effects:
photon scattering off dust in 30 Doradus, continuum/line nebular
emission, pixel saturation and stellar crowding. Dust scattering and
gas emission raise the background flux and, in turn, make it more
difficult to resolve stars. The net result is visible in the diffuse
``nebulous'' areas of the completeness map (see Figure
\ref{map50}). More localized than the dust and gas effects, pixel
saturation produces the visible ``streaks'' (charge-bleeding along the
rows of the CCD) extending off of many bright stars. Finally, the
extreme crowding conditions near R~136 are responsible for the central
completeness ``hole'' (where most of the injected fake stars are
lost).

\subsection{Reddening and data-model comparison}

We are not assuming any a priori reddening distribution. In principle
we do not know where populations of different age are located
\emph{inside} NGC~2070, so reddening is an unknown function of
age. Naively speaking, young massive stars may have carved the gas
with their ionizing flux, hence lowering the extinction. On the other
hand, high levels of SF could be sustained only where the gas density
is higher, which would lead to the opposite situation. Besides,
normalization and reddening of the reference field are also free
parameters. Furthermore, PMS stars may appear reddened due to their
local circumstellar material.

When dealing with such a large parameter space (age, reddening and
field contamination) there is uncertainty on whether the best solution
is local or global. To cope with this we combined a genetic algorithm
with a local search procedure (Hybrid-Genetic Algorithm, HGA; see
Appendix \ref{app} for details) which is more effective to avoid local
trapping than local search alone. In our approach the first step is to
store a library of ``basic'' synthetic CMDs, where each CMD is a Monte
Carlo synthetic population generated with a step-wise SF and reddening
distribution.

When the CMDs are generated other population parameters like
metallicity, distance modulus (m$-$M)$_{0}$, slope of the IMF and
binary fraction q are kept fixed at 0.008 (e.g., \citealt{luck98}),
18.5 (e.g., \citealt{panagia91,schaefer08,pietr13}), \cite{kroupa01}
and 30\%\footnote{Primary and secondary stars are both picked from the
  same IMF.}, respectively. The extinction coefficients $A_{\lambda}$
are taken from \cite{demarchi14b}. The explored ages range from now up
to 50 Myr ago. Since older isochrones tend to be more tightly packed
in the CMD than younger isochrones, the duration of each age step
increases with age from 1 Myr (at the present time) to 20 Myr (50 Myr
ago). The reddening is allowed to vary between 0 and 1 mag with a step
of 0.05 mag.

\begin{figure*}[!t]
\centering \includegraphics[width=15cm]{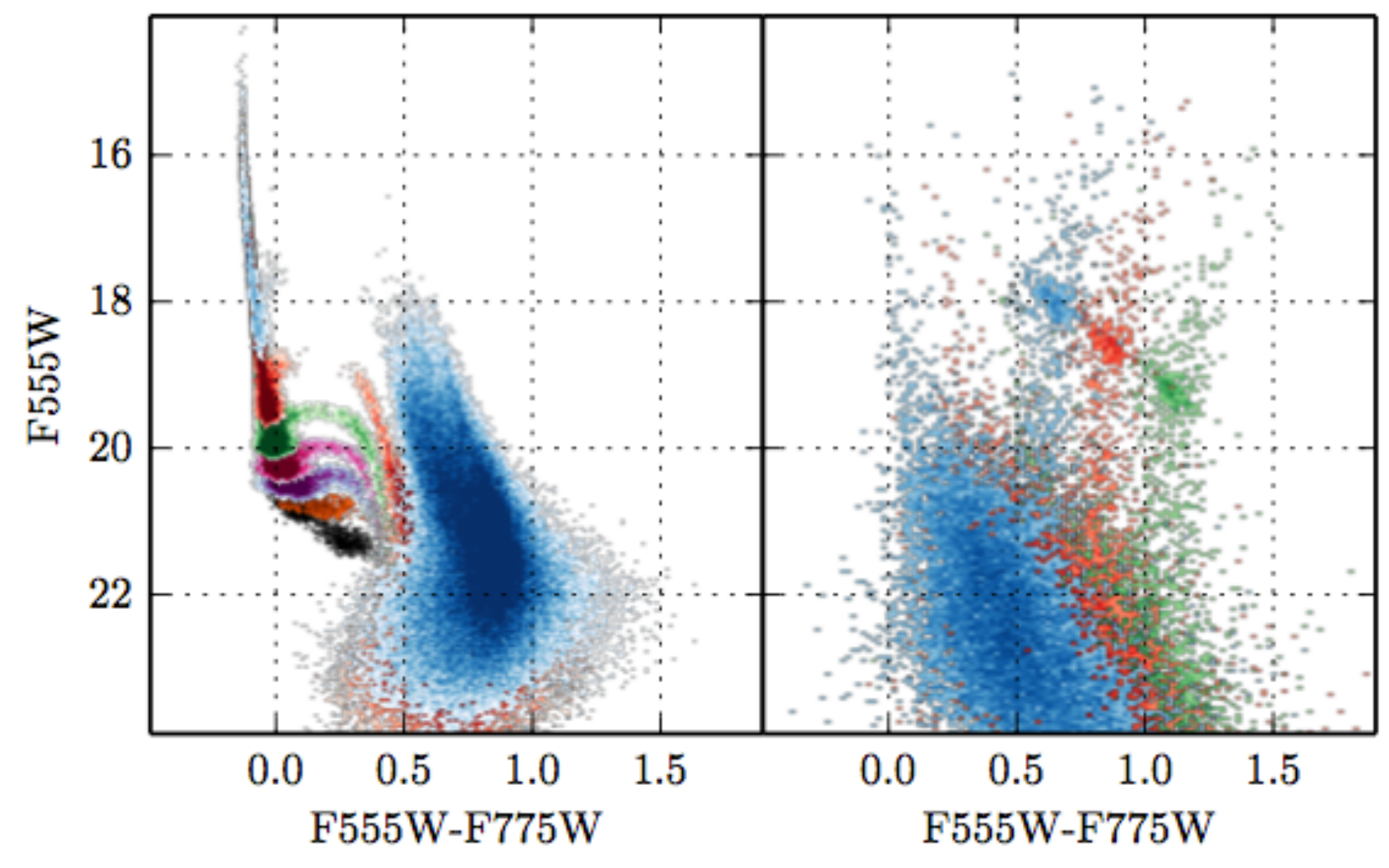}
\caption{Left panel: example of ``basic'' synthetic CMDs color-coded
  with age: blue, red, green, purple, violet, orange and black stars
  have ages in the range 0-1 Myr, 2-3 Myr, 4-5 Myr, 6-7 Myr, 8-9 Myr,
  10-12 Myr, 14-16 Myr, 18-20 Myr respectively; Right panel: example
  of ``basic'' field CMDs color-coded with reddening: blue, red, green
  stars are field stars artificially reddened with E(B$-$V) in the range
  0-0.05 mag, 0.35-0.40 mag and 0.70-0.75 mag respectively.}
\label{basic_cmds} 
\end{figure*}

All ``basic'' synthetic CMDs (see left panel of Fig. \ref{basic_cmds})
are then degraded using the photometric errors and incompleteness as
derived from the artificial star tests discussed in the previous
Section. Once this basis is generated, any complex synthetic CMD can
be constructed as a linear combination of the ``basic'' synthetic
CMDs. In order to take into account field contamination, the CMD of
the reference field (see Section \ref{rc_sec}) is artificially reddened
with steps of 0.05 mag between E(B$-$V)=0 and E(B$-$V)=1 to produce a
complete basis of field CMDs. Likewise any kind of contamination can
be simulated by linearly combining these ``basic'' field CMDs (see
right panel of Fig. \ref{basic_cmds}). The only difference is that the
``basic'' field CMDs are corrected for the \emph{difference} in
photometric errors and incompleteness as estimated in the field itself
and in the NGC~2070 region, while the ``basic'' synthetic CMDs are
only corrected for the latter. However, since in the reference field
completeness and photometric errors at F555W$\sim 24$ are 100\% and
$1/20$ of the error in sub-region A1, the first correction is almost
negligible.
The combination of ``basic'' CMDs (both synthetic and field) which
minimizes the residuals from the observational CMD (in terms of
Poissonian likelihood) is searched with the HGA code. The best
coefficients tell us the most likely: 1) star formation rate as a
function of time (i.e., the SFH); 2) total reddening (foreground $+$
internal) as a function of time; and 3) field contamination (reddening
and normalization).

\begin{figure*}[!t]
\centering \includegraphics[width=9cm]{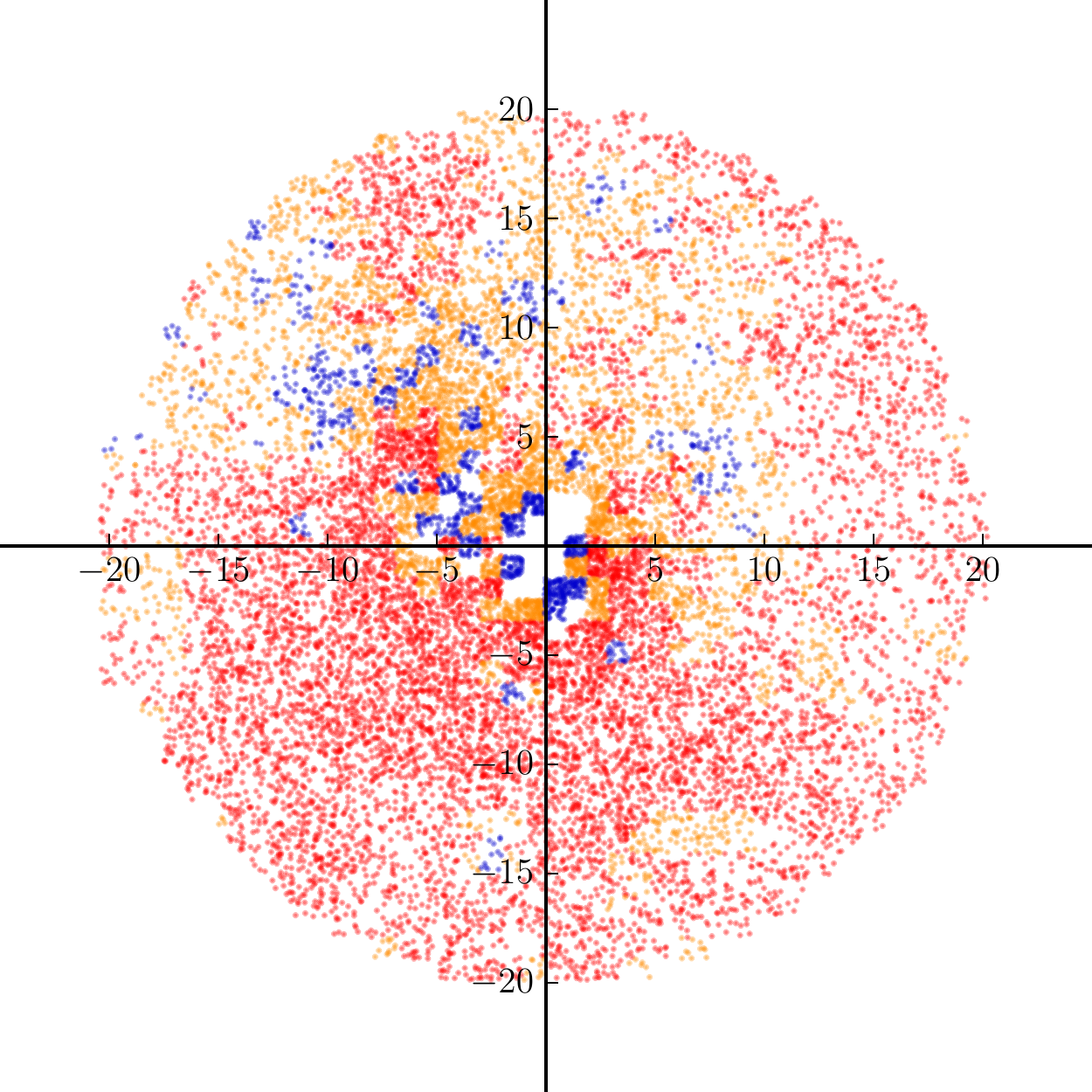}
\caption{Sub-regions of NGC~2070 where the average completeness is
  50\% for F555W $> 24$ (A1; red dots), $23<$F555W$<24$ (A2; orange
  dots) and $22<$F555W$<23$ (A3; blue dots).}
\label{map3a} 
\end{figure*}

\emph{Which part of the CMD is used?} Not all of the CMD is used to
recover the SFH. Our analysis is limited on the bright end by the
saturation magnitude and on the faint end by the 50\% completeness
magnitude level. Nonetheless, even with these conservative selections,
the large variations of completeness with magnitude in some regions
(see upper plot in Fig. \ref{map50}) force us to derive the SFH in
sub-regions within which the completeness is reasonably uniform. In
fact, our artificial stars are uniformly distributed across the
region, whereas real stars are concentrated in clumps and filaments,
structures covering a minor fraction of the total area. As a
consequence, most of the stars will suffer a real incompleteness which
is worse than the average measured with artificial stars. The division
in ``iso-complete'' sub-regions mitigates this bias. Fig. \ref{map3a}
shows our selection: sub-regions of NGC~2070 where the completeness in
F555W is 50\% for F555W $> 24$ (A1), $23<$F555W$<24$ (A2) and
$22<$F555W$<23$ (A3) are indicated in red, orange and blue,
respectively. The stars found in both the F555W and F775W filters
inside these sub-regions are used to recover the optical
SFHs. Overall, our sub-regions trace the stellar density, from the
lowest (A1) to the highest (A3).

This analysis leaves out the very center of NGC~2070, a 3 pc region
mostly represented by the SSC R~136. Here stellar crowding is so severe
that most of the intermediate and low mass stars are lost, hence
little, if any, information is available from TOns.

To further validate the SFH, we independently analized the NIR
data. To do this we used all stars detected in both the F110W and
F160W images inside each sub-region (defined using the F555W data). In
this case however the faint limiting magnitude was anchored to the
50\% completeness level in F110W.

\section{Results}

\subsection{SFHs}
\label{sfhs}
\begin{figure*}[!t]
\centering \includegraphics[width=12cm]{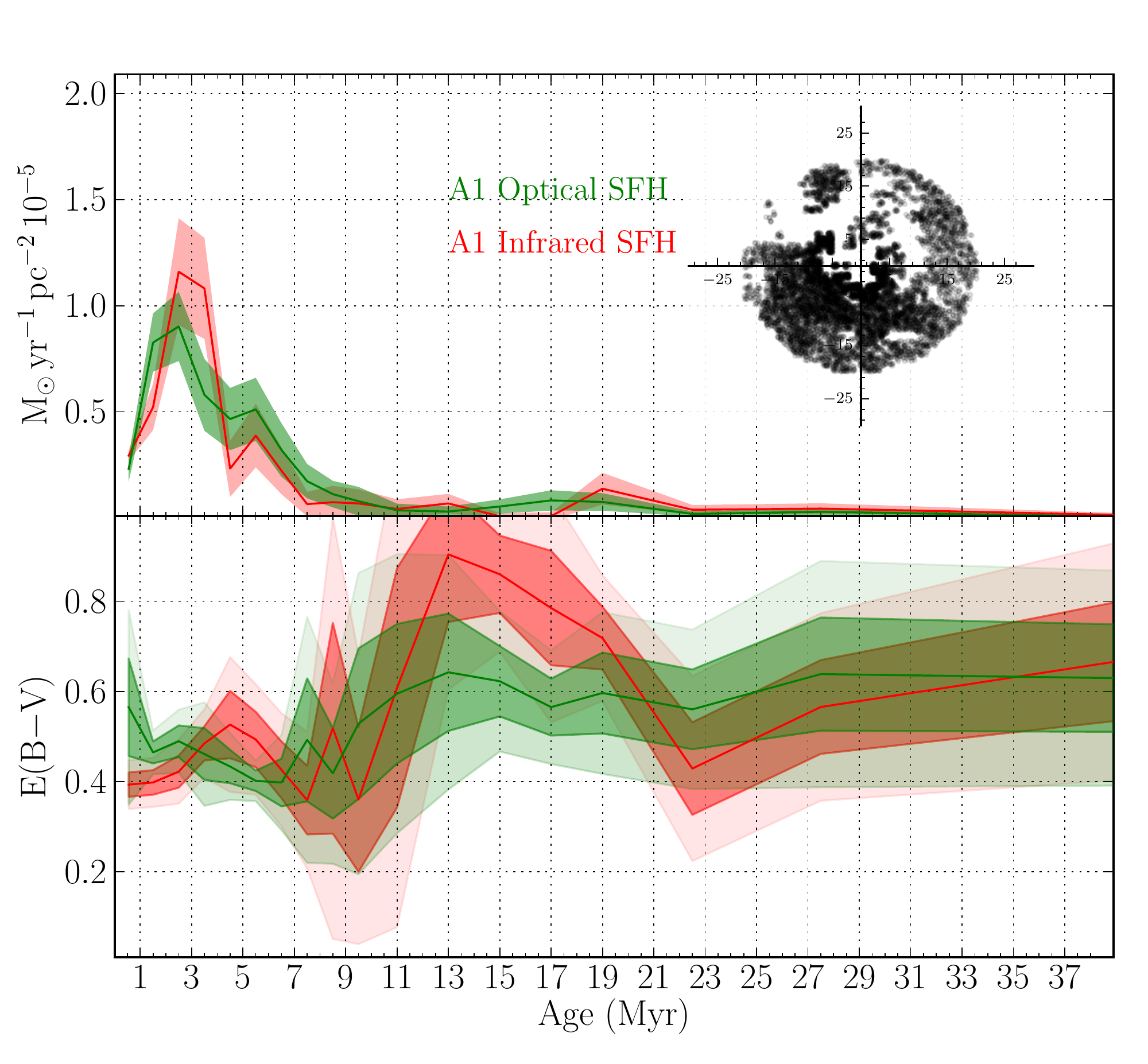}
\caption{Recovered SFH (top panel) and reddening distribution (bottom
  panel) for region A1. The optical and NIR solution are plotted in
  green and red colors, respectively. The inset panel shows the
  distribution of stars in region A1.}
\label{sfra1} 
\end{figure*}

\begin{figure*}[!t]
\centering \includegraphics[width=12cm]{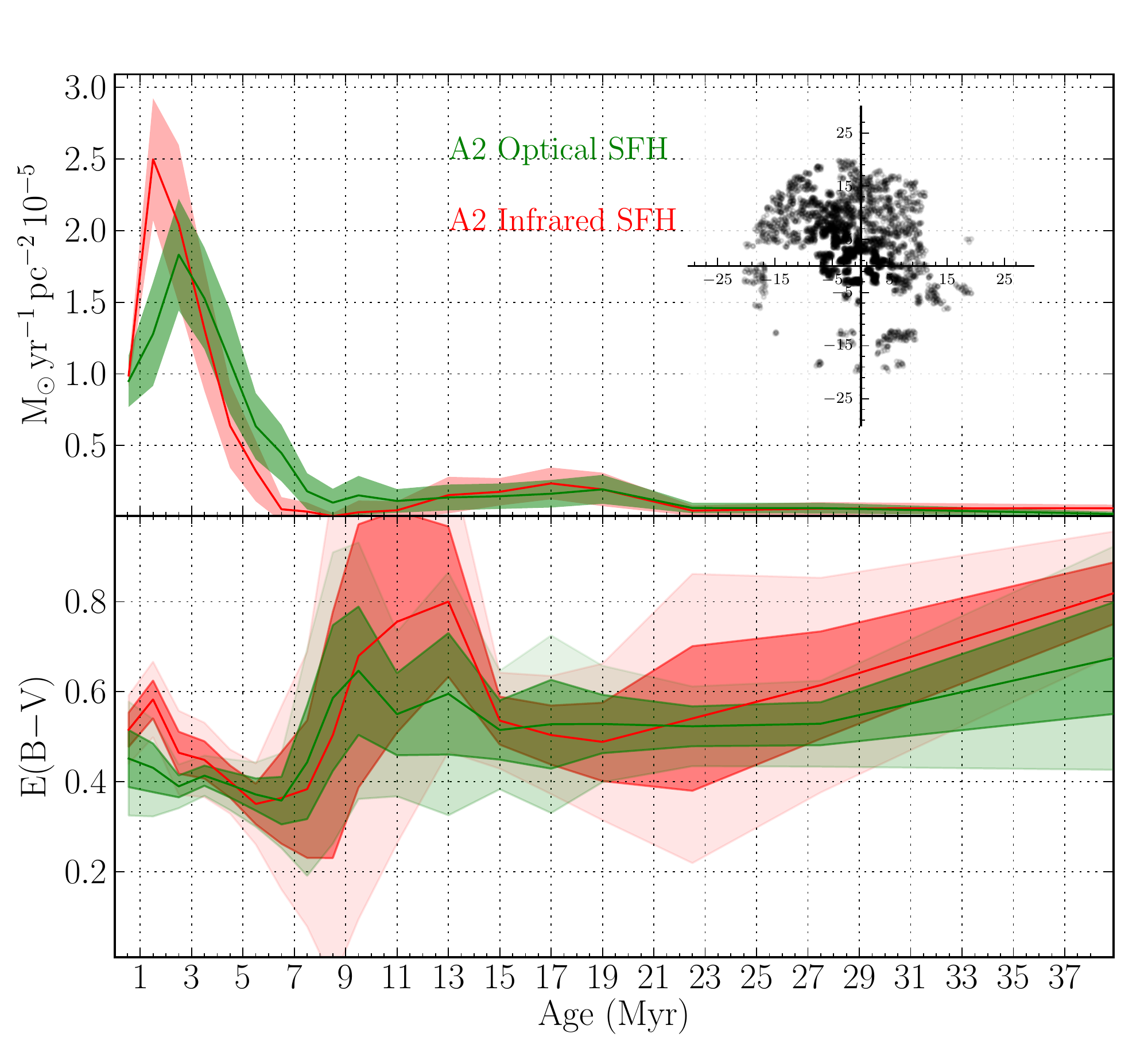}
\caption{Same as Fig. \ref{sfra1} but for region A2.}
\label{sfra2} 
\end{figure*}

Figures \ref{sfra1} and \ref{sfra2} show the recovered SF rate (in
$\mathrm{M}_{\odot}\,\mathrm{yr}^{-1}\,\mathrm{pc}^{-2}$; top panel)
and reddening as a function of age (bottom panel) for sub-regions A1
and A2 respectively, as predicted using the optical (green curve) and
NIR (red curve) CMDs. In principle our code provides the full
distribution of reddening as a function of age, but to ease the
visualization we only provide the mass-weighted\footnote{Weighting
  with the predicted mass allows us to take into account that high
  reddening stars tend to be under-sampled because of the
  incompleteness.} average reddening as a function of age. The shaded
bands indicate one (darker) and two (lighter) standard
deviations. Such uncertainties are the quadrature sum of a statistical
error (obtained by bootstrapping the data and re-deriving the
solutions) and systematic error (obtained by re-deriving the solutions
with different age-binnings and CMD binning scheme).

Despite the different spatial resolution and reddening sensitivity,
optical and NIR predictions are in good agreement. Both results
predict that mild SF activity started throughout the whole NGC~2070
region $\approx 20$ Myr ago, and that about 7-8 Myr ago the birth rate
accelerated. Interestingly, the activity in the last 1 Myr has been
relatively quiet compared to the average in the last 5 Myr.

Concerning the activity that commenced 20 Myr ago, it is important to
note that our SFHs have been obtained by taking into account field
contamination, hence, unless we have been very unlucky in our field
selection\footnote{Normalization and reddening of field stars are let
  to vary, whereas the ratio between \emph{young and old field stars}
  depends on the chosen field.}, it is reasonable to assume that this
activity is a local extras over the LMC field. Nonetheless, whatever
the origin, its significance is low (zero activity is still within
1-sigma error bars).

\begin{figure*}[!t]
\centering \includegraphics[width=8cm]{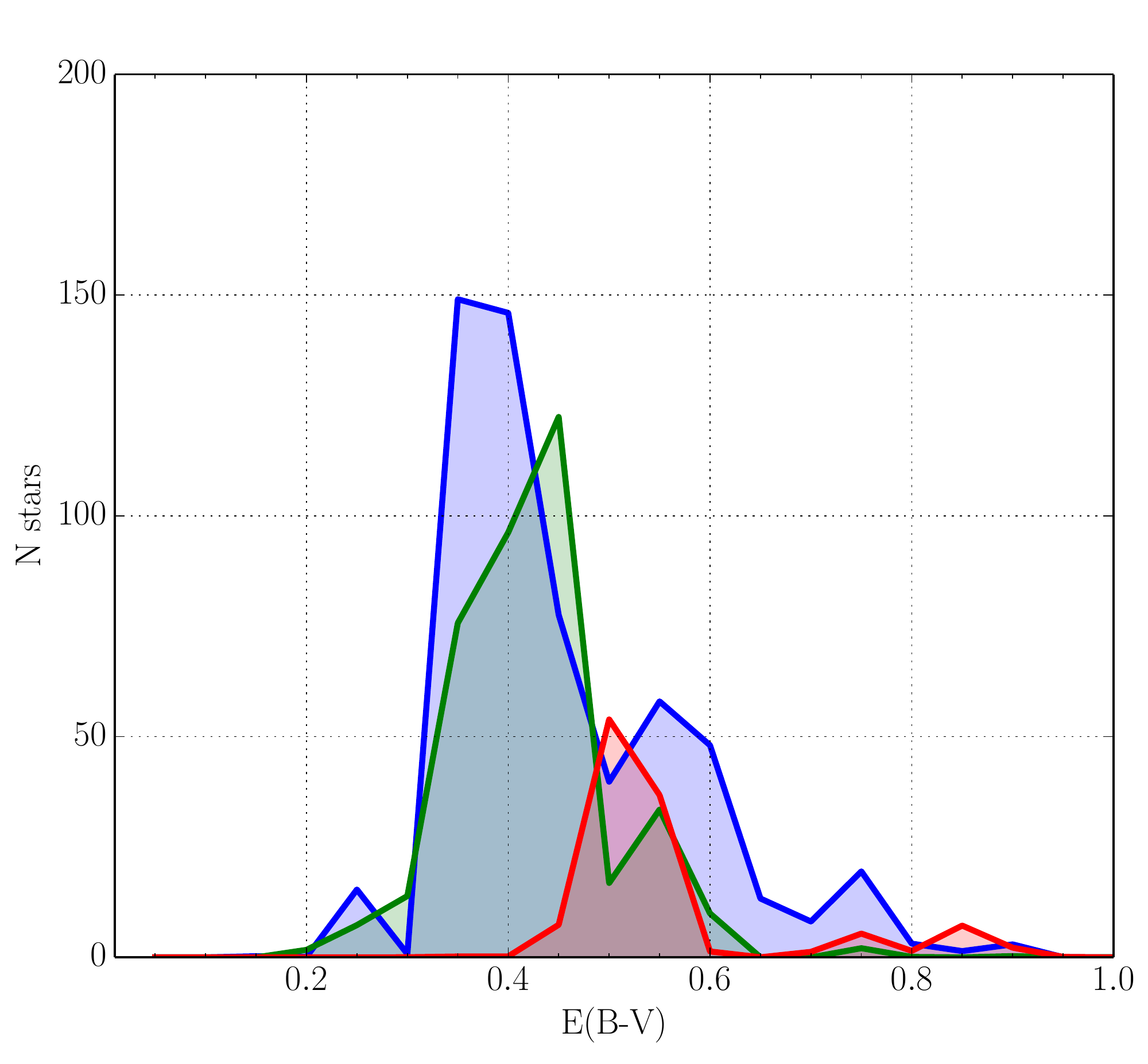}
\caption{Predicted number of stars in region A1 as a function of
  reddening E(B$-$V) for three age bins, 2 Myr (blue), 6 Myr (green)
  and 20 Myr (red).}
\label{full_red} 
\end{figure*}

Focusing on the specific sub-regions, the SFH of region A1 shows a
mild bimodality, with a minor peak around 5-7 Myr ago and a major peak
1-4 Myr ago (1-3 Myr ago in the optical SFH, 2-4 Myr ago in the NIR
one). The measured reddening distribution anti-correlates with SF: the
average E(B$-$V) is $\approx 0.6$ for stars with ages older than 7
Myr, when the SF activity was lower, and $\approx 0.4$ for stars of
younger ages, when the SF was stronger. Optical and NIR reddening
derivations agree within errors, both in confirming the presence of
notable differential reddening. The large oscillations of the NIR
solution are mostly due to the low sensitivity to reddening in these
filters. Figure \ref{full_red} shows an example of the full reddening
solution (number of considered stars vs E(B$-$V)) for region A1 in
three age bins (2, 6 and 20 Myr). As it can be seen, the full
distributions are asymmetric with a long tail at high
reddening.

The SF in sub-region A2 (Fig. \ref{sfra2}) is about two times higher
than in sub-region A1. The most important feature in the A2 SF is a
pronounced peak in both optical and NIR solutions 1-3 Myr ago. Like in
sub-region A1, the reddening of sub-region A2 is anti-correlated with
the SF activity. The average reddening values are also very similar.

\begin{figure*}[!t]
\centering \includegraphics[width=12cm]{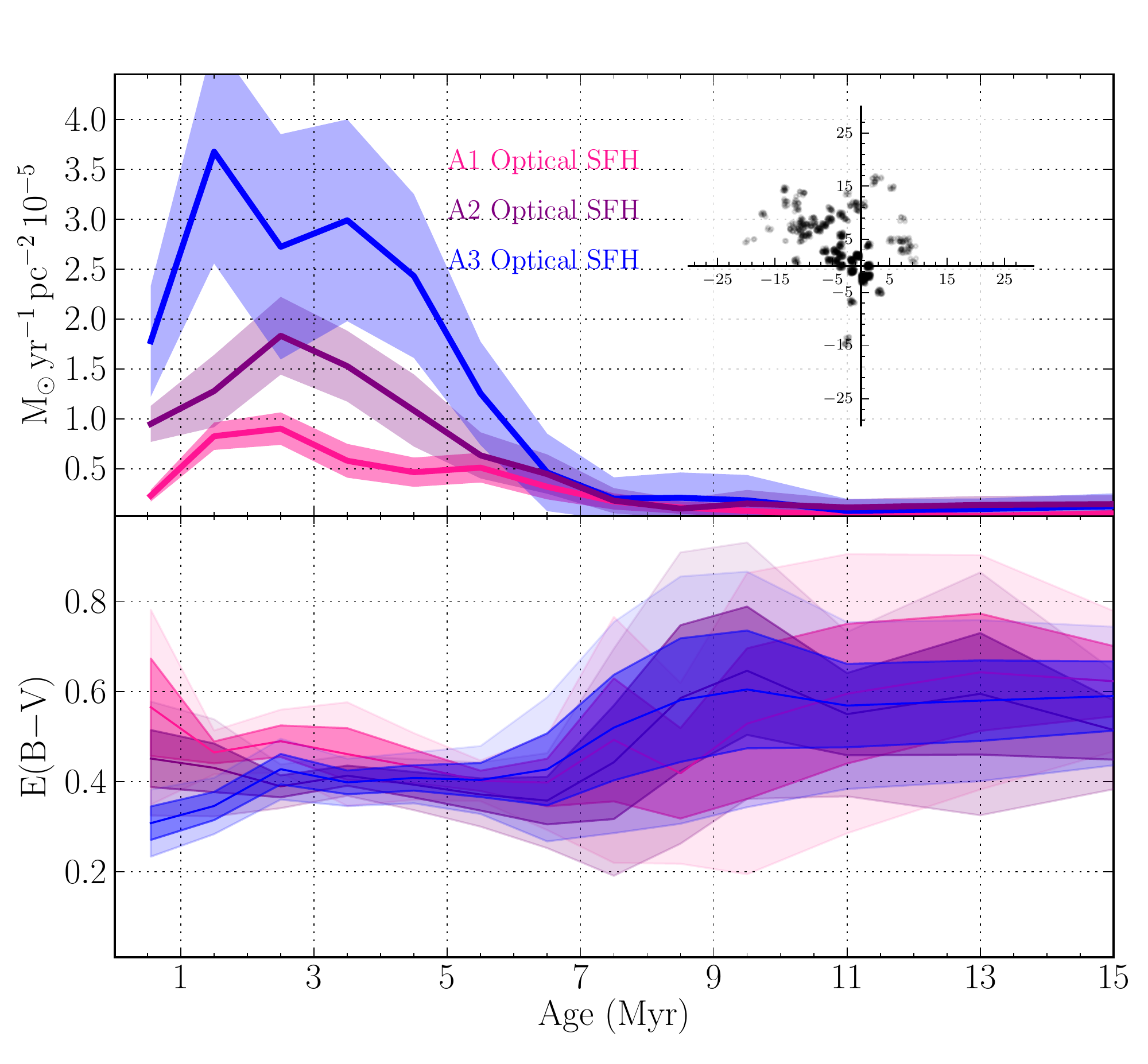}
\caption{Optical SFHs (top panel) and reddening solutions (bottom
  panel) for regions A3 (blue), A2 (purple) and A1 (pink). The inset
  panel shows the distribution of stars in region A3.}
\label{sfra3} 
\end{figure*}

Because of the high crowding we used only the optical CMD to derive
the SFH of the most internal sub-region A3. Fig. \ref{sfra3} shows the
result (blue) compared to the other two SFHs over the last 15 Myr. The
bottom panel shows the corresponding reddening distribution. The peak
activity is almost three times higher than in A2, and at odds with
other sub-regions, is clearly located between 1 and 2 Myr ago.

The total mass assembled in the regions A1, A2, and A3, in the last 7
Myr, is about $2.9\,\times 10^{4}\,\mathrm{M}_{\odot}$, $2.5\,\times
10^{4}\,\mathrm{M}_{\odot}$, and $1.1\,\times
10^{4}\,\mathrm{M}_{\odot}$, respectively.

Finally, our analysis does not involve the central core of NGC~2070,
where the activity is presumably much higher. A rough estimate can be
made by re-scaling the SFH of region A3 to match the number of stars
in the magnitude range F555W 16-18, where the sample is fairly
complete even in the core. In this magnitude range the core contains
two times the stars of region A3, but concentrated in an area four
times smaller. This leads to a scaled peak rate of the order of $2.8\,
\times 10^{-4}\,\mathrm{M}_{\odot}\,
\mathrm{yr}^{-1}\,\mathrm{pc}^{-2}$ and a total mass of $2.2\,\times
10^{4}\,\mathrm{M}_{\odot}$, in excellent agreement with
\possessivecite{hunter95} estimate. Adding this mass estimate to the
mass of sub-regions A1, A2 and A3, we get a total mass for NGC~2070 of
$8.7\,\times 10^{4}\,\mathrm{M}_{\odot}$\footnote{We stress that this
  result has been obtained with a \cite{kroupa01} IMF. Changing from a
  Kroupa to a Salpeter IMF below 0.5 $\mathrm{M}_{\odot}$ increases
  the total by a factor 1.6.}. This value is compatible with the
results of \cite{selman99}, who found $5.5\,\times
10^{4}\,\mathrm{M}_{\odot}$ within 14 pc from R~136 (using a Salpeter
IMF down to 0.5 M$_{\odot}$), \cite{andersen09}, who found
$2.7\,\times 10^{5}\,\mathrm{M}_{\odot}$ (using a Salpeter IMF down to
0.1 M$_{\odot}$) and \cite{bosch01}, who found
$10^{5}\,\mathrm{M}_{\odot}$ using dynamical considerations.

\subsection{Fit quality}

\begin{figure*}[!t]
\centering \includegraphics[width=14cm]{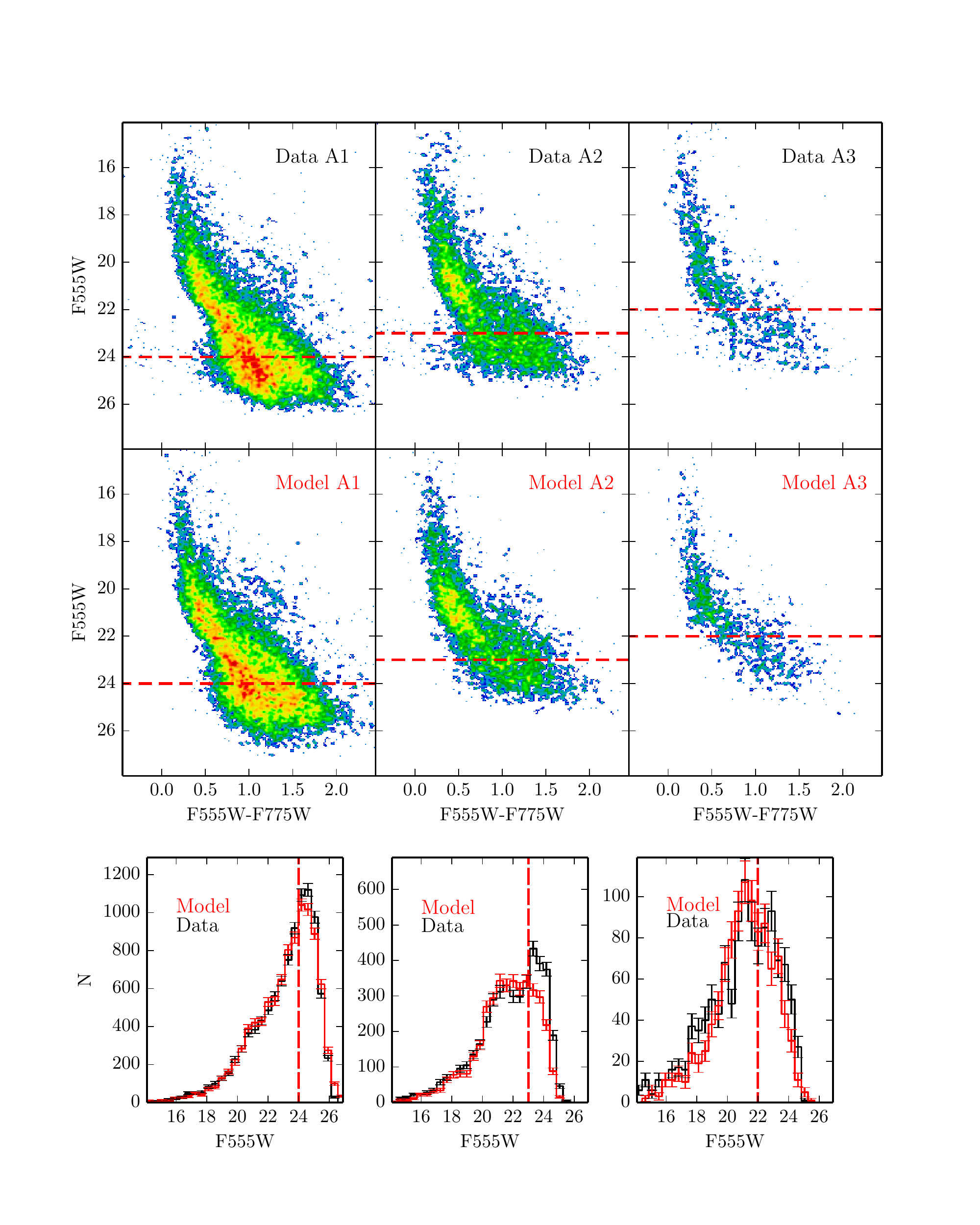}
\caption{Observational optical CMDs (top panels) and best synthetic
  CMDs (middle panels) for sub-regions A1, A2 and A3 (from left to
  right). Bottom panel: data vs model LFs. The dashed line indicates
  the magnitude limit used to recover the SFH.}
\label{cmd_synth_vi} 
\end{figure*}

\begin{figure*}[!t]
\centering \includegraphics[width=10cm]{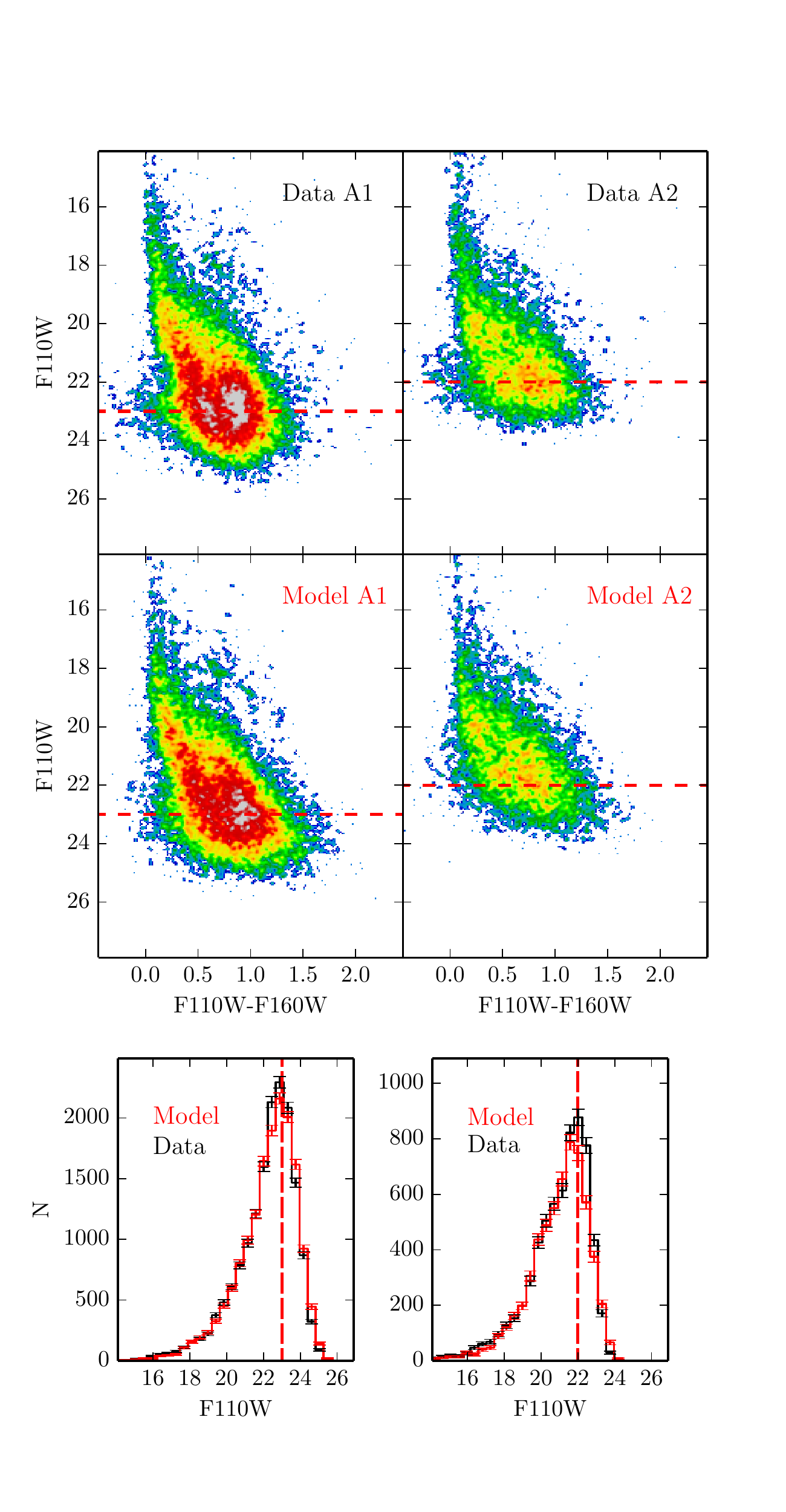}
\caption{Same as Fig. \ref{cmd_synth_vi} but for NIR data and
  sub-regions A1 and A2 only. The dashed line indicates the magnitude
  limit used to recover the SFH.}
\label{cmd_synth_jh} 
\end{figure*}

Figure \ref{cmd_synth_vi} compares the observed optical CMDs (top
panels) in the three sub-regions (from left to right, A1-A2-A3) to
synthetic CMDs (middle panels) generated from our best solutions. The
bottom panels show the corresponding LFs. The red dashed line
corresponds to the faintest limit used to fit the data.  Figure
\ref{cmd_synth_jh} shows the same analysis for the NIR CMDs (regions
A1 and A2 only).

Simulations for sub-regions A1 and A2 show a good agreement with
both the optical and NIR CMDs. Within the portion of the CMDs used to
fit the data (see red dashed lines), observational and model LFs show
deviations that are compatible to the errors (computed as square root of the
observational count rates). This suggests that a Kroupa IMF, which is
very similar to a Salpeter IMF above 0.5 M$_{\odot}$, is consistent
with the data. At fainter magnitudes our simulations systematically
underestimate the observational star counts (more in the optical than
the NIR). This effect is minor in sub-region A1, but becomes
significant in sub-region A2. One possible explanation for this
mismatch is the 0.75 mag selection threshold that we used to reject
artificial stars that are actually blended with real stars. Relaxing
this condition alleviates the issue, but also increases the scatter in
the synthetic CMDs more than what we see in the data. However, the impact
of this effect should be minor in the magnitude range used for
fitting.

The simulation for sub-region A3 is in less good agreement with the
observations. As is visible from the LFs (see the bottom right panel
of Fig. \ref{cmd_synth_vi}), the observed star counts systematically
outnumber synthetic counts even at bright magnitudes (F555Ws$< 20$).
The straightforward explanation is to attribute the discrepancy to the
incompleteness of the region, which is too inhomogeneous to be
accurately modeled by our tests. However, from a physical point of
view, it is also conceivable that the transition from PMS to MS for
intermediate/massive stars is not well reproduced by models. Indeed,
the major discrepancy is in the range F555W 18-19, which corresponds
to TOn ages younger than 1 Myr. If the PMS evolutionary time for these
objects is overestimated (so more PMS stars were predicted than
observed) or the PMS birth-line (the region in the CMD where stars are
still embedded in their gas cocoons and, therefore, still invisible in
optical bands) is closer to the MS than predicted, any attempt to
simultaneously fit PMS and MS star counts will end up with a bias. In
this case, the deficiency of MS stars could suggest that the SF rate
for stars younger than 1 Myr might be underestimated.

Finally, the number and colors of RC stars are well reproduced. This
indicates that field contamination and reddening are correctly
modeled.

\subsection{Solution robustness}

\begin{figure*}[!t]
\centering \includegraphics[width=14cm]{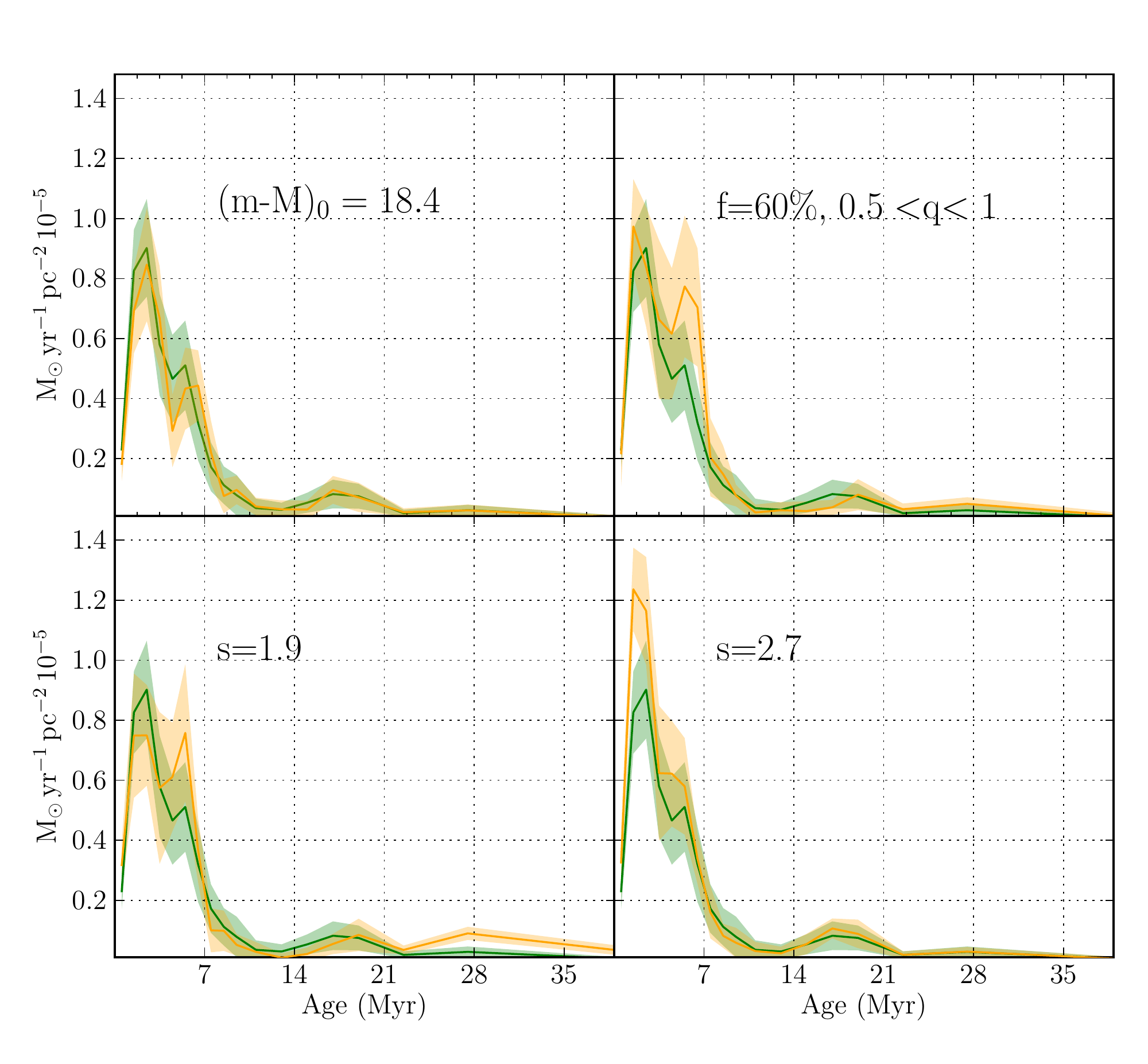}
\caption{Sensitivity test for the SFH in sub-region A1. Solutions for
  different assumptions for the IMF, binary population and distance
  (orange shaded histograms) are overlaid on the standard solution
  (green shaded histogram) . The top-left and top-right panels show
  the result of changing distance and binary fraction from 18.5 and
  30\% (primary and secondary masses randomly paired from the same
  IMF) to 18.4 and 60\% (mass ratio randomly drawn from a constant
  distribution between 0.5 and 1), respectively. The bottom-left and
  bottom-right show the result of changing the IMF exponent above
  $1\,\mathrm{M}_{\odot}$ from 2.3 to 1.9 and 2.7 respectively.}
\label{imf_bin} 
\end{figure*}

To test the robustness of the solution against variations in the
assumed IMF, binary prescription (fraction and mass ratio) and
distance, we re-derived the SFH of region A1 using alternative values
as an example. In Figure \ref{imf_bin} we compare the results (shaded
orange histograms) with the standard solution (shaded green
histogram). The top-left panel shows the SFH obtained using a shorter
distance modulus, (m$-$M)$_{0}=18.4$, as suggested by some Cepheids
studies (e.g. \citealt{macri06}). The top-right panel shows the SFH
obtained with a binary fraction of 60\% (e.g. \citealt{sana13}) and
mass ratio randomly drawn from a constant distribution between 0.5 and
1. The bottom-left and bottom-right panels show the SFH obtained using
an IMF exponent $s$, above $1\,\mathrm{M}_{\odot}$, of 1.9 and 2.7,
respectively.

Overall, all recovered solutions are qualitatively and quantitatively
(within 2-sigma) consistent, suggesting that our findings are
robust. In particular, the onset of the major activity 7 Myr ago is
unchanged. Relatively larger differences are found for IMF
changes. The predicted rate for s$=1.9$ is significantly stronger than
the $s=2.3$ case for ages $>20$ Myr, while for s$=2.7$ it is stronger
for ages $<5$ Myr. These differences are principally due to the SF
rate-IMF degeneracy. The lack of stars of lower mass (caused by the
flatter IMF) is compensated with a higher early SF rate, whereas the
lack of intermediate and massive stars (caused by the steeper IMF) is
compensated with a higher recent activity.

We also found that the synthetic CMDs corresponding to the $s=1.9$ and
$s=2.7$ solutions systematically underestimate and overestimate,
respectively, the star-counts for F555W$<20$ (the reverse for
F555W$>23$). On the other hand, changing the binary fraction and mass
ratio does not affect either the SFH or the fit quality.

\section{Comparison with Previous Studies}
\subsection{SFH}
\cite{selman99} recovered the SFH of NGC~2070 using a Bayesian
approach applied to the UBV photometry down to $V=19.2$, corresponding
to stars more massive than $20\,\mathrm{M}_{\odot}$. The SFH was found to be
dominated by three episodes, namely a young peak at $0 < t < 1.5$ Myr,
an intermediate-age peak at $1.5 < t < 3.5$ Myr, and an old peak $4 <
t < 6$ Myr ago. These three bursts appear to be spatially disjoint,
with the youngest stars concentrated towards the center while the
intermediate-age stars appear spherically distributed over a 6 pc
radius, slightly off-center. The observations are consistent with a SF
that propagated inward.

\cite{andersen09} obtained HST/NICMOS F160W band images of the central
14 pc $\times$ 14.25 pc around R~136 and combined them with archival
WFPC2 F555W and F814W observations (\citealt{hunter95,hunter96}). By
fitting the F160W vs F555W$-$F160W CMD with stellar models above
$7\,\mathrm{M}_{\odot}$ \citep{marigo08} and stellar models below
$7\,\mathrm{M}_{\odot}$ \citep{siess2000}, they constrained the age of the low
mass population to be $2-4$ Myr old.

\cite{brandl96} derived the age distribution for a region of $13\times
13$ arcseconds$^2$, at a distance of 4 arcseconds from R~136, using a
synergy of NIR photometry with adaptive optics and HST
photometry. Their investigation was limited to stars more massive than
12 M$_\odot$. The resulting age distribution (see their Fig. 13) grows
from 7 Myr ago up to the present time with three visible bursts at 5-6
Myr, 3-4 Myr and 1 Myr ago.  Moreover, they did not find red giants
and red super-giants in their FOV, concluding that the age of R~136
must be less than 5 Myr.

According to \cite{walborn97} the central region can be divided into a
core, R~136, $2-3$ Myr old, a peripheral triggered population $<$ 1 Myr
old, and a group of late-O and early-B stars 4-5 Myr old.

\cite{demarchi11} used deep HST observations including H${\alpha}$
for a field of $\sim 3$ arcmin $\times\,3$ arcmin enclosing
NGC~2070. They inferred that a significant fraction ($\sim$ 35\%) of
the PMS stars were formed prior to 12 Myr ago, while a similar
fraction is younger than 4 Myr. In terms of SF, they found that 1 Myr
ago the region was about 30 times more active than 16 Myr ago.

Using isochrone fitting, \cite{sabbi12} found that the majority of the
stars in the ``northeast clump'', a group of stars a few pc away from
R~136, have ages between 2 and 5 Myr ago, while stars in R~136 are at
most 2 Myr old.

Except for \cite{demarchi11} and \cite{sabbi12}, our data are
generally much deeper than the others in the literature and,
therefore, more sensitive to older activity. Indeed, our prediction
about the beginning of activity in NGC~2070, about 20 Myr ago, is only
accessible with our data. Moreover, our analysis implements the PARSEC
evolutionary models, which cover consistently the entire evolution
from the PMS phase to the post-MS phase, whereas all previous analyses
had forced to combine models for the PMS phase (\citealt{siess2000,
  tognelli11}) and other models for later phases.

Despite these differences, our SFH is broadly consistent with the
aforementioned results. Of particular interest is the 7 Myr epoch,
when our solution predicts a significant SF enhancement. This event is
generally consistent, or slightly older, with what is found in all
other studies. Similarly to \cite{demarchi11} we find evidence of some
activity prior to 10 Myr ago, whose rate is, at most, one order of
magnitude lower than the recent one.

From a spatial point of view, the activity shifts to younger ages
moving from our region A1 to A3, resembling the inward progression
found by \citealt{selman99}. This is interesting, given that
\citealt{selman99} and this work use complementary mass ranges, above
$20\,\mathrm{M}_{\odot}$ in the case of \citealt{selman99} and below
$6\,\mathrm{M}_{\odot}$ in our case. The inward scenario is also
broadly consistent with the \cite{demarchi14b}, \cite{walborn97} and
\cite{sabbi12} findings. \cite{demarchi14b} found that stars younger
than 4 Myr are more concentrated towards R~136 than stars older than
12 Myr. These authors also found a remarkable lack of old activity
($>\,12$ Myr) near R~136, whereas the old component is not
significantly different in our sub-regions. However, the data in our
sub-regions A2 and A3 are not deep enough to allow for a conclusive
argument.

\subsection{Reddening}

Our results predict that stars younger and older than about 10 Myr are
on average reddened by E(B$-$V)$\sim 0.4\pm 0.05$ and $\sim 0.6\pm
0.1$, respectively. Because young stars are spatially more concentrated,
this also translates into a negative reddening gradient towards the
center. This trend is in apparent contrast with findings of
\cite{zar99}, who found that the young populations (star forming
regions few Myr old) in the LMC are more reddened than the older ones
($>$ 1 Gyr). However, we point out that the timescales here are much
shorter, our ``old'' population is only few Myr older the young
one. Besides, all results indicate a large reddening dispersion (at
least 0.1 mag).

Looking at the literature, we find a general consensus for highly
variable reddening in the NGC~2070 neighborhood. More specifically,
\cite{demarchi14b} used UMS and RC stars to study extinction around
NGC~2070. Once corrected for the foreground contribution
(E(B$-$V)=0.07), our average prediction for the young component is
very close to the peak of their reddening distribution obtained from
UMS stars (see the blue histogram in their Fig. 9), while our average
prediction for the old component is well within their reddening
dispersion.

Our reddening for the young population is also in good agreement with
the results of \cite{selman99}. These authors used stars more massive
than $20\,\mathrm{M}_{\odot}$ and found average extinctions in the range
$A_{V}=1.1-1.5$ (E(B$-$V)=0.35-0.5).

\section{Comparison with other starburst clusters}

Overall, NGC~2070's SFH is similar to that in NGC~346, the largest
star forming region in the Small Magellanic Cloud (SMC), whose major
activity started about 6-8 Myr ago and peaked about 3 Myr ago before
dropping to a lower level (see \citealt{cignoni11}). If we exclude the
core of NGC~2070, which is ten times more active than NGC~346, even
the peak rate of region A3 is similar in amplitude to the peak rate in
NGC346 (about $2\times 10^{-5}\,\mathrm{M}_{\odot}\,
\mathrm{yr}^{-1}\, \mathrm{pc}^{-2}$ between 4 and 5 Myr ago;
\citealt{cignoni11}).

\cite{niede} analysed the CMDs of eight young massive LMC clusters and
derived upper limits on potential age spreads by fitting Gaussian
profiles to their SFHs. They found age spreads smaller than a few Myr
for the youngest clusters (20 -- 60 Myr old), which is consistent with
our much more detailed analysis for NGC 2070.

Outside the Local Group, resolving older populations in SSCs becomes
very challenging. \cite{larsen11} analysed seven young massive star
clusters (5 -- 50 Myr old) in five nearby galaxies, located at
distances of 3 -- 5 Mpc. They found that the simulated CMDs generated
with Padova isochrones \citep{bertelli09} and single burst SFH have
problems to reproduce: 1) the observed separation in the CMDs between
MS and He-burning ``blue loop'' stars, 2) the ratio between red and
blue supergiants and, in some clusters, 3) the scatter in the
luminosities of the supergiant stars (scatter that cannot be explained
by observational errors alone). The authors could improve the fit by
including an age spread of 10--30 Myr in model clusters.

\section{Conclusions}

We have presented a detailed analysis of the star formation history in
the starburst cluster NGC~2070, located in the heart of 30 Doradus in
the LMC, using deep optical and NIR CMDs from the Hubble Tarantula
Treasury Project. We used a new synthetic CMD approach combined with
the latest Padova models (PARSEC), the first to cover homogeneously
all stellar phases from PMS to post-MS.\\

In our implementation we encountered a number of interesting
challenges. We summarize here our main conclusions:\\

\emph{SFH:} we found that NGC~2070 experienced prolonged activity,
starting at least 7 Myr ago. We identify three major events
in the history of this cluster:

\begin{itemize}

\item $\approx 20$ Myr ago - This epoch demarcates the commencement of
  the first significant period of SF. Prior to this epoch, local
  activity is not distinguishable from the average activity in the LMC
  field;
\item 7 Myr ago - The SF accelerated throughout the entire region;

\item 1-3 Myr ago the activity reached a peak. In this time range, the
  SF moves from the periphery to the central regions. Our innermost
  region (A3) shows a maximum activity 1-2 Myr ago;
\end{itemize}

\emph{Stellar mass:} We estimate the stellar mass of NGC~2070 out of
20 pc to be $\approx 8.7\,\times 10^{4}\,\mathrm{M}_{\odot}$. This
value is close to the median mass of Galactic globular clusters
($8.1\times 10^{4} \mathrm{M}_{\odot}$; \citealt{mandu91}).

\emph{Reddening:} Concerning the reddening distribution, we find an
average E(B$-$V)$\approx 0.4$ mag for the young population ($<$10 Myr
old), and $\approx 0.6$ mag for the old one. An explanation could be
that only in the last few Myr the SF has been vigorous enough to sweep
away part of the gas through stellar winds. Another possibility is
that the old activity took place on the far side of the NGC~2070
nebula as seen from us, therefore those stars experience most of the
line of sight optical depth.

\emph{IMF:} Except the innermost few pc, where the incompleteness is
too severe to allow firm conclusions, a Kroupa IMF down to
$0.5\,\mathrm{M}_{\odot}$ is compatible with the data. This
corroborates and extends the result of \cite{andersen09} to sub-solar
masses. To the level we can measure low-mass stars can form in
starburst clusters in the same way they form in low density
environments.

\acknowledgments We would like to thank Mario Gennaro, Nino Panagia,
Martha Boyer, Chris Evans, Jay Gallagher, Karl Gordon, Anton
Koekemoer, Soren Larsen and Selma de Mink for helpful comments,
discussions, and contributions to other aspects of
HTTP. E.K.G. gratefully acknowledges funding via the Collaborative
Research Center ``The Milky Way System'' (SFB881) of the German
Research Foundation (DFG), particularly via subproject B5. M.T. has
been partially funded by the Italian PRIN-MIUR grant
2010LY5N2T. D.A.G. kindly acknowledges financial support by the German
Research Foundation (DFG) through grant GO\,1659/3-2. Support for
program \#12939 was provided by NASA to the US team members through a
grant from the Space Telescope Science Institute, which is operated by
AURA, Inc., under NASA contract NAS 5-26555.

\appendix

As done previously (\citealt{cignoni06,cignoni11}), we parametrize the
synthetic CMD as a linear superposition of basic CMDs generated with
step-function-like star formation, metallicity and reddening. In the
specific case of this paper we deal with 19 steps in age and 20 steps
in reddening, for a total of 380 parameters. To explore this wide
parameter space we have combined a well tested genetic algorithm (GA),
Pikaia\footnote{Routine developed at the High Altitude Observatory,
  and available in the public domain
  http://www.hao.ucar.edu/public/research/si/pikaia/pikaia.html.},
with a local search routine. As shown in various papers (see, e.g.,
\citealt{ng02}, \citealt{aparicio09}, \citealt{small13}), GAs allow to
find global optimum more efficiently than a local search alone. In our
approach the synergy of the GA and a local search combines the
advantages of both worlds. In the next sections we describe the
synthetic population code and the optimization routine (GA$+$local
search). Finally, the capabilities of the approach are tested with
artificial data.\\

\section{Synthetic CMDs}

\label{app}

The basic synthetic CMDs$(j,k)$ are populated through the following
Monte Carlo procedure: 1) synthetic masses and ages are extracted from
the assumed IMF and the j-th SF step, respectively; and 2) synthetic
masses and ages are converted to absolute synthetic magnitudes and
colors by using a fine grid of isochrones. For our calculations we
used the latest (V.1.2S) PARSEC isochrones, covering the entire mass
spectrum of $0.1-350\,\mathrm{M}_{\odot}$ from the PMS phase to the early-AGB
phase; 3) A fraction q of synthetic stars is randomly chosen to have a
companion. The masses of companions are extracted from the same IMF
and their flux is added to the flux of the primaries; 4) The absolute
synthetic photometry is put at the distance of the LMC and reddened
with the k-th reddening. To produce realistic simulations, all basic
CMDs are degraded with photometric errors and incompleteness as
estimated from artificial star tests (see Section \ref{artifix}).

To have all stellar phases well populated, the synthetic CMDs are
generated with a large number of stars ($10^6$). Once constructed, the
basic CMDs$(j,k)$ are binned in n bins of color and m bins of
magnitude. The final result is a library of j$\times$k 2D histograms
CMD$_{m,n}(j,k)$ and any CMD can be expressed as a linear combination
of these 2D histograms (see Equation \ref{eq1}). The coefficients
$S(j,k)$ that multiply each of the CMD$_{m,n}(j,k)$ are the star
formation rate at the time step j and reddening step k. The sum over j
and k of $\mathrm{S}(j,k)\times \mathrm{CMD}_{m,n}(j,k)$ provides the
total star-counts predicted $\mathrm{N}_{m,n}$ in the CMD bin (m,n) by
the star formation $S(j,k)$.
\begin{eqnarray}
\label{eq1}
\mathrm{N}_{m,n}= \sum_{\substack{j,k}} [\mathrm{S}(j,k)\times \mathrm{CMD}_{m,n}(j,k)]
\end{eqnarray}
Including the reference field in the models corresponds to changing
Equation \ref{eq1} into Equation \ref{eq2}:
\begin{eqnarray}
\label{eq2}
\mathrm{N}_{m,n}= \sum_{\substack{j,k}} [\mathrm{S}(j,k)\times
  \mathrm{CMD}_{m,n}(j,k)]+\\+S_{F}(k)\times F_{m,n}(k)]\nonumber
\end{eqnarray}
where S$_{F}$(k) regulates the number of field stars with reddening
k-th, while F$_{m,n}(k)$ is the actual number of stars in the reference
field (reddened with the k-th reddening) in the CMD bin (m,n).

\section{Best solution search}

Once the observational CMD is binned as well, the next step is to
search for the combination of basic CMDs that minimizes the CMD
residuals between data and model. For this task we implemented a
likelihood distance, whose minimization is not biased by low count
statistics. The combination of basic CMDs that minimizes the
likelihood corresponds to the most likely SFH behind the data. The
uncertainty around the recovered best solution will be the sum in
quadrature of a statistical error, obtained through a data bootstrap,
and a systematic error, obtained by re-deriving the SFH using
different age-binnings and CMD-binning.

 In our approach the likelihood is minimized with an hybrid-genetic
 algorithm (HGA), which combines a classical GA with a local
 search. Pure GAs are iterative probabilistic algorithms for solving a
 problem that mimic processes found in the natural biological
 evolution. Compared to local search algorithms, GAs explore the
 search space in more points simultaneously, hence they are far less
 sensitive to the initial conditions and show remarkable ability to
 escape from local minima. Shortcomings of GAs are the weak ability
 for local exploration and the slow convergence rate. The proposed HGA
 aims to overcome both by alternating two phases: a GA, whose goal is
 to search for a quasi-global solution, and a local search, whose goal
 is to increase solution accuracy. The synergy of the two incorporates
 the exploration ability of GAs and the exploitation ability of local
 search algorithm. For this study we implemented both \emph{parallel}
 and \emph{serial} hybridization. The parallel one is applied to each
 iteration and aims at enhancing offspring likelihood by means of a
 local search before moving to the next generation. The serial one
 consists in applying a final local search after which the GA
 population has evolved to the region containing the global solution.

\section{Test}
\begin{figure*}[!t]
\centering \includegraphics[width=10cm]{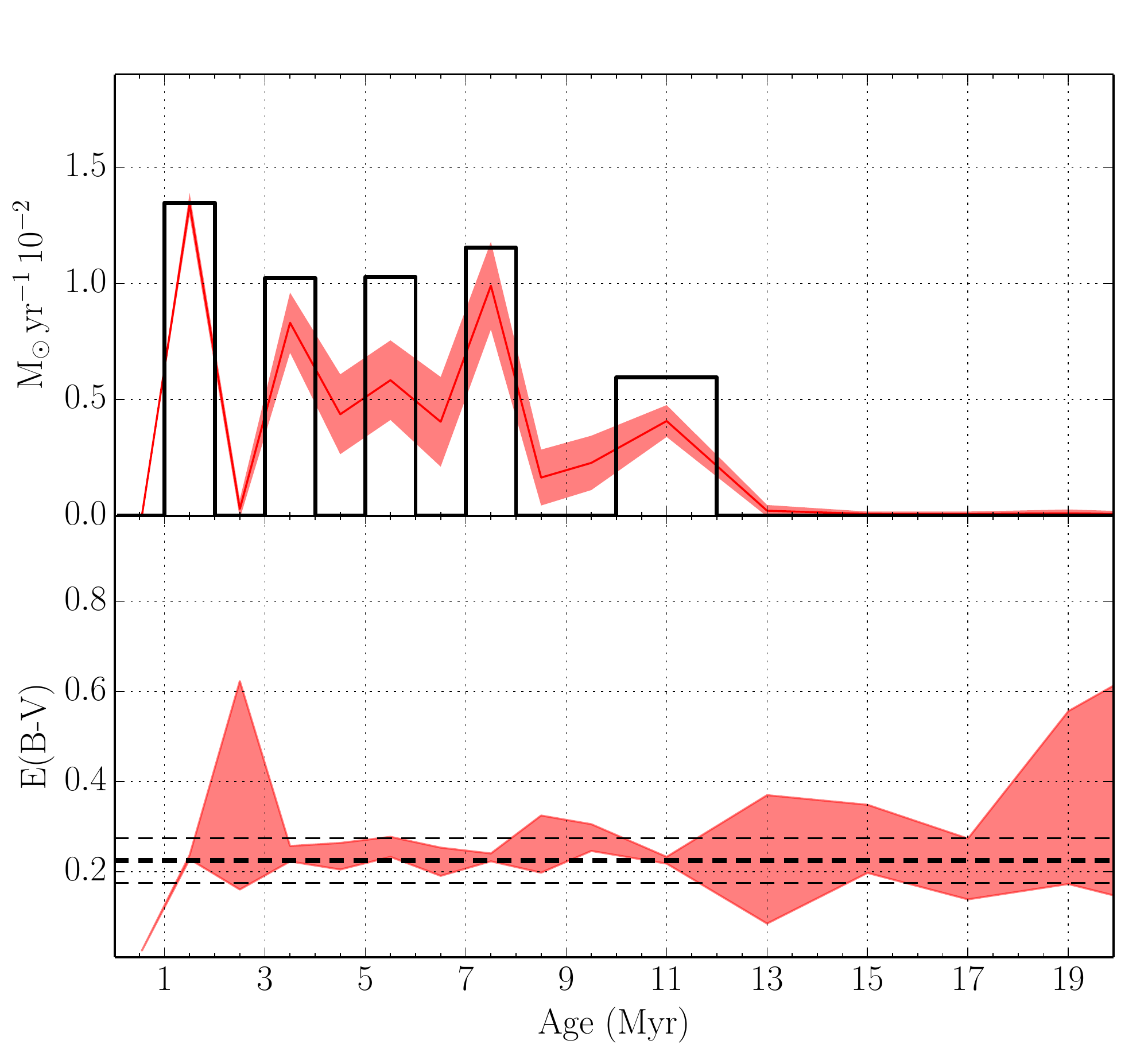}
\caption{SFH recovery test. The top panel shows the SFH (input in
  black, recovered in red), bottom panel show the E(B$-$V) (input in
  black, recovered in red).}
\label{test} 
\end{figure*}

To test our approach we have generated synthetic data with features
resembling NGC~2070 (10000 stars brighter than F555W$=24$). To
simulate the observational conditions, this fake population is
convolved with photometric uncertainties and incompleteness from the
artificial star tests of NGC~2070. The input SFH is a sequence of five
bursts of different ages and duration 1 Myr, while the input reddening
distribution is built with 50\% of the stars with E(B$-$V) between
0.20 and 0.25, and 25\% between 0.25 and 0.30, 25\% between 0.15 and
0.20. Figure \ref{test} shows the result of the reconstruction. The
input SFH and reddening distribution are in black; the best
reconstructed counterparts in red. The recovered SFH looks like a
smoothed version of the input SFH, due to the finite time resolution
associated with the discrete nature of the data, and the
ill-conditioned nature of mathematical inverse problems. However,
overall, input functions are successfully recovered, with most of the
differences within the uncertainties. Moreover, a bursty SFH is
recovered more accurately if the time gap between the individual
bursts is longer.


\end{document}